\documentclass[%
preprint,
onecolumn,
titlepage,
tightenlines,
12pt
]
{revtex4-2}

\usepackage{subcaption}

\usepackage{dcolumn}
\usepackage{cancel}
\usepackage{tikz,pgfplots}

\usepackage{amsmath, bm, amssymb,graphicx}

\usepackage[export]{adjustbox}

\begin{document}

\preprint{APS/123-QED}

\title{Dynamic interactions between an aerodynamic flow and a flexible
flat plate}

\author{Srikumar Balasubramanian}
 \email{sb44@illinois.edu}
\affiliation{%
 Department of Aerospace Engineering, University of Illinois at Urbana-Champaign, IL 61801, USA
}%

\author{Andres Goza}%
\affiliation{%
 Department of Aerospace Engineering, University of Illinois at Urbana-Champaign, IL 61801, USA
}%


\begin{abstract}
Passive flow control via fluid-structure interaction (FSI) is a promising paradigm for unmanned aerial vehicles operating in vortex-dominated low Reynolds number regimes. The associated aerodynamic flows are unsteady, high-dimensional and nonlinear in nature. A flexible structure therefore has the potential to passively alter key unsteady vortex structures through its vibrations, if its intrinsic modal dynamics are carefully aligned with the driving flow processes. Towards this aim, we perform high-fidelity numerical FSI simulations to study the dynamic interplay between a separated aerodynamic flow with vortex-shedding content ($Re=500$ and angle of attack of $15^{\circ}$) and a  flexible flat plate modeled using linear Euler-Bernoulli beam theory. Our choice of low Reynolds number, $Re = 500$, is relevant to small-scale unmanned aerial vehicles and has received less attention in literature, enabling insights into the coupled dynamics for a less studied flow regime. To isolate the natural evolution/instability mechanisms of the coupled dynamics, we start our simulations from a steady base state of the coupled FSI system. The ensuing dynamics involve a departure from the base state to an attractor that has two clear distinctions from the dynamics of rigid plate-FSI: the existence of a mean structural deformation (yielding a camber effect) and dynamics about this mean deflected state. The structural dynamics about the mean camber can be represented through a small subset of structural modes, with relative importance to the overall structural response dictated by the alignment of the structural natural frequency to the vortex-shedding frequency. The effect of the mean camber on the key flow phenomena is well studied in literature, but there remain open questions about the coupled dynamics that we focus on in this work: (i) do the early-time dynamics that evolve from the base state involve a coupled or flow-driven instability?; (ii) how are the flow dynamics and associated lift informed by the modal shape(s) and deformation timescales of the structural response, and how is this structural response informed by the structural parameters relative to the vortex-shedding behavior of the reference (rigid) case? For question (i), we identify that the departure from the base state arises from a flow-induced instability, suggesting that it is the vortex shedding of the aerodynamic flow that instigates structural motion, which eventually yields coupled dynamics that vary based on the structural parameters. For question (ii), we further clarify how the structural mode dynamics in the FSI setting are dictated by the relationship between the vacuum fundamental frequencies of the structure and the baseline rigid-plate shedding frequency. We show that the long-time FSI dynamics can be partitioned into distinct regimes categorized by the dominant mode of the structural dynamics. The coupled dynamics in each regime are described in detail, highlighting in particular the effect of the dynamic interplay between the structure and flow and drawing connections to the baseline rigid case as appropriate. 
\end{abstract}
\maketitle 
\section{Introduction}\label{sec:intro}
Efficient maneuver of the next generation of unmanned aerial vehicles (UAVs) requires sophisticated flow control strategies. Flow control techniques can be categorized as active and passive depending on the mechanism used in effecting control \cite{gad-el-hak_2000, activecontrolreview}. Active strategies involve momentum/energy injection through external actuators (e.g., plasma jet injection \cite{jetreview,jetcontrol1}, blowing \cite{blowing_review}, zero-net-mass-flux devices \cite{znmf}) to alter the flow features. On the other hand, traditional passive techniques achieve flow alterations using static obstructions/modifications to the flow-structure interface (examples include using turbulators, surface roughness, surface cavities, vortex generators; see \citet{passivereview} for a review). While the active and passive techniques described above have achieved considerable success, there remain challenges to their full integration into aerodynamic flow control strategies. Active strategies often involve complex and heavy actuator components, and the feedback control laws for their operation involve sophisticated reduced order models that are still in development \cite{rom_review_1, rom_review_2}. Static techniques cannot adapt to unsteady flow dynamics and typically yield detriments at off-design conditions.  

Compliant surfaces offer a compelling paradigm by which passive (and therefore low energy), adaptive actuation can be obtained without the need for complex feedback laws or actuator designs. The resulting surface oscillations have the capacity to affect surrounding flow structures \cite{kang2015, lei2014}, and the interface motion is driven by the surrounding flow, avoiding the need for heavy actuators. Coupled flow-structure interactions  could lead to the excitation of specific modal responses in the structure, that could in-turn lead to modifications to key flow phenomena \cite{li_jaiman_khoo_2021, jaiman2022}. 

Prior work has investigated the mechanisms of interplay between compliant structures and aerodynamic flows. \citet{gordnier} performed high-fidelity numerical simulations to study the fluid-structure interaction (FSI) between a membrane wing airfoil and aerodynamic flow at 
 a Reynolds number of $2500$. The ensuing FSI was categorized into two broad categories: effects due to the static mean structural deformation and the dynamics associated with the unsteady structural motion. The effects of angle of attack and structural parameters on each category were studied separately. \citet{song2008} experimentally studied the interactions between a compliant membrane and incoming flow at a Reynolds number range of $7\times10^4-2\times10^5$. It was found that having the mean static positive structural deformation led to an increase in both the mean lift coefficient and the lift-to-drag ratio. The curved positive shape of the structure delayed stall and ensured that the flow was attached even at large angles of attack. Other numerical and experimental studies \cite{li_jaiman_khoo_2021,staticeffect,Gehrke_2022} have compared the static effects of compliant wing-FSI with that of rigid cambered wings. These comparison studies have enabled a better understanding of the fluid-structure interaction effects associated with the induced mean-camber.

While the static FSI effects are well understood in literature, the coupled interactions associated with the dynamic structural oscillations and key aerodynamic flow phenomena remain less well characterized. Of particular interest here are how the coupled dynamics are selected based on an alignment between the vortex-shedding timescales in the flow and the intrinsic structural vibration timescales (i.e., natural frequencies, which might need to account for flow contributions depending on how large the structural parameters are relative to their associated flow quantities). \citet{xi2020} experimentally studied the FSI of a membrane wing at a fixed angle of attack of $\alpha=14^{\circ}$, for a Reynolds numbers of approximately $6\times10^4$. From the study, the dominant frequency associated with the coupled dynamics was found to be either the structural vibration frequency or the dominant baseline shedding frequency. Interactions were studied for three different Reynolds numbers and it was reported that the frequency of the coupled dynamics  matched the structural natural frequency in one case, and matched the baseline vortex-shedding frequency in the other two cases. \citet{xu2017} performed a numerical study at a range of Reynolds numbers ($100-10,000$) to highlight the effect of structural flexibility in dictating the vortex shedding-structural vibration interactions. The excited structural modes were identified through phase portraits and bifurcation diagrams for a fixed angle of attack and Reynolds number. In the same vein, \citet{ganapathisubramani2017} performed an experimental study at a Reynolds number of $56{,}000$, where the coupled interactions were characterized by the phase relationship between the flow and structural dynamics. In-phase dynamics between flow and structure were found to be beneficial for aerodynamic performance. The experimental study of \citet{Rojratsirikul2010} at Reynolds numbers of $Re= 53{,}100,$ $79{,}700$, and $106{,}000$ identified that the structural oscillations, determined by the structural and flow parameters, could excite shear layers and reduce the separation regions. The study focused on how the excited structural modes varied as a function of angle of attack and incoming flow velocity. \citet{GORDNIER2014} performed numerical simulations to study membrane-vortex interactions at $Re=24{,}300$ and three different angles of attack $\alpha= 10^{\circ},$ $16^{\circ},$ and $23^{\circ}$. The membrane's mode one oscillations were found to significantly reduce the size of the separation region and enhance lift. 

Many studies have reported how a flexible wing in aerodynamic flow could undergo dynamics with a range of observed structural modes.  \citet{song2008}  hypothesized that the the structural mode selection mechanism could be due to forcing from the flow-induced instability; i.e, relative proximity between vortex-shedding frequency and the structural natural frequencies. \citet{jaiman2022} performed numerical simulations of membrane-wing FSI at a Reynolds number of $24{,}300$ and angles of attack of $15^{\circ}$, $20^{\circ}$, and $25^{\circ}$. The study identified the dominant flow-structure mode that is excited at each angle of attack, and probed the cause behind the ``lock-on" that occurred between unsteady vortex-shedding and the structural vibrations. The proposed mechanism was a feedback cycle that involved a structural mode selection and subsequent interaction with the vortex shedding.  \citet{li_jaiman_khoo_2021} studied membrane-wing FSI at $Re = O(10^3-10^5)$, for an angle of attack of $15^{\circ}$ and a variety of structural mass and stiffness parameters. The coupled interactions were found to depend on the relation between the vortex-shedding frequency and the fundamental frequencies associated with the membrane.

Motivated by this prior work, we aim to answer the following questions. First, is the origin of the fluid-structure interplay driven by a flow or a coupled FSI instability? \citet{li_jaiman_khoo_2021} have reported the onset of the coupled dynamics as being flow excited, but a systematic assessment of this question would benefit from a study where the dynamics are initiated from a formal equilibrium state of the fully coupled system. From this initial condition, characterizing the onset of dynamics for a variety of flow and structural parameters would help identify mechanistic origins for the coupled dynamics.

Second, for the long-time dynamics, what are the various regimes possible by tuning the flow and structural parameters to systematically vary the relationship between the inherent vortex-shedding frequency of a rigid reference case and the natural frequencies of the structural system? \citet{li_jaiman_khoo_2021} identified that the alignment of these frequencies is key to selecting the FSI dynamics, and a parametric sweep informed by the relative values of these frequencies (rather than a sweep in terms of the raw structural parameters agnostic to relative frequency considerations) can clarify in more detail what physical processes separate the behavioral regimes, and for which regimes effective structural parameters from the flow play a role. Of particular focus for this study is a clearer assessment of (i) the boundaries in parameter space that delineate when the various structural modes are selected, and what role that fluid-induced mass has in that selection process; (ii) what categorically distinguishes the long-time coupled FSI dynamics for each selected structural mode, and what trends in time-averaged lift arise from these qualitatively distinct sets of behavior; (iii) within and across the various identified FSI regimes, what the detailed interplay is between the structural motion and the vortex shedding structures that are formed, how these change from one another and from the baseline (rigid) case, and which sets of changes are beneficial or detrimental to lift.

Third, the majority of these studies have been at higher Reynolds numbers, and we therefore focus here on the above mentioned questions of this FSI system at a lower Reynolds number of $Re=500$ relevant to small-scale aerial vehicles and during take-off and landing of larger unmanned aerial craft. 

We note that the vast majority of studies for this flow-compliant aerodynamic body system have involved a membrane body, which has no bending stiffness and therefore the FSI dynamics arise as a result of the dynamic balance between fluid loads with internal in-plane structural stresses and structural inertia. We focus here, by  contrast, on a linear Euler Bernoulli beam because of its connection to more rigid craft that have potential utility in the aforementioned unmanned aerial craft. These beam structures produce only out of plane stresses in responses to stimuli. That said, the coupled dynamics arise as a balance between the fluid loads, the structural inertia, the (out-of-plane versus in-plane) internal structural stresses, and the inherent timescales that govern this exchange. We consider only linear structural models here because of the small structural displacements (the results below show that the majority of displacements are significantly less than 10\% of the plate length) and because the linear representation allows for an unambiguous definition of structural natural frequency, which facilitates our study that is rooted in contrasting this structural natural frequency with the underlying vortex-shedding process of the rigid-body system.

To address our targeted questions, we perform high-fidelity numerical simulations. To address the first question of the origin of the FSI dynamics, we begin our numerical FSI simulations from the formal base-state of the flow-structure configuration and demonstrate that the origin of the ensuing instability is flow-driven and induced by the bluffness of the plate to the flow. Starting from the base state also enables us to explore the evolution from this linear flow-induced instability to the long-time attractors that arise for different structural parameters.  To answer the second question related to delineation of the key regimes, we incorporate the aforementioned linear Euler-Bernoulli beam model into a flow at a Reynolds number of $Re=500$ and an angle of attack of $15^{\circ}$, for which a rigid flat plate exhibits vortex shedding. In this way, the structural parameters can be varied to understand the effect of the relative structure natural frequency to the vortex-shedding frequency. For these systematically chosen structural parameters (informed a priori by the targeted frequency relationship), we study both the early-time FSI evolution from the base-state as well as the long-time dynamics. Regimes are identified that are associated with qualitatively distinct behavior and structural dynamics of predominately mode one, mode two, and mode three based on the relationship between the modal frequency's proximity to the shedding behavior. Within the structural mode one, there is a sub-regime that is entirely driven by the effective structural properties imposed by the flow---changes of the structural parameters do very little to change the dynamics in that regime. Moreover, within each regime, the interplay between the structural dynamics and the formation and advection of the vortex-shedding process is studied in detail. The circulation, spatial wavenumber, and timescales of key vortex structures are quantified within each behavioral regime, connected to the spatio-temporal dynamics of the structures, and used to explain changes in aerodynamic lift.

\section{Problem setup and numerical scheme}\label{sec:setup}
\subsection{Problem setup}
\begin{figure}[hbt!]
\centering
 \input{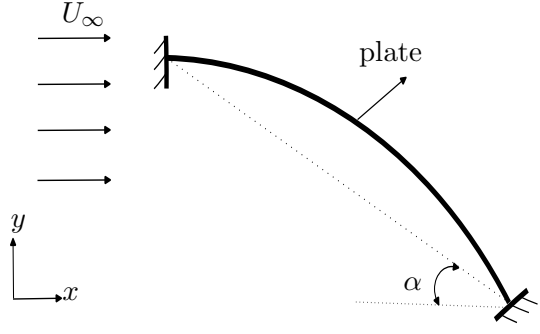}
\caption{Schematic of the flow past a flexible Euler Bernoulli beam configuration used in this study.}
\label{fig:schem}
\end{figure}

A schematic of the problem under consideration is shown in figure \ref{fig:schem}. The incoming flow $U_{\infty}$ is from left to right and the  Reynolds number ($Re$) based on the length of the plate ($l$) is $Re=\frac{U_{\infty}l}{\nu}$, where $\nu$ is the kinematic viscosity of the fluid. The flexible flat plate is inclined at a sufficiently large angle of attack, $\alpha=15^{\circ}$, to induce flow separation at the leading and trailing edges, and an associated periodic vortex-shedding process. Our interest is how the fluid-structure interplay alters the nominal vortex shedding associated with a rigid flat plate. We use linear Euler-Bernoulli beam theory to model the flexible flat plate. The Euler-Bernoulli structure is constrained using clamped-clamped boundary conditions. In this study, we vary the mass and bending stiffness of the plate. The structural parameters may be non-dimensionalized in terms of the flow parameters as 
\begin{equation}
    m = \frac{\rho_sh}{\rho_f l}, \; k= \frac{E_{s}I}{\rho_{f}U_{\infty}^2l^3h}, 
    \label{eqn:params}
\end{equation}
 where $m$ is the structure-fluid mass ratio and $k$ is the structure-fluid stiffness ratio, respectively. The above mentioned dimensionless parameters are commonly used in the broader FSI literature, (see, e.g., references \cite{connell_yue_2007,shoele_mittal_2016,gurugubelli_jaiman_2015,goza2017} for some representative studies). Here, $\rho_{s}$ is the density of the plate and $\rho_{f}$ is that of the fluid; $E_{s}$ is the Young's modulus of the plate and $I$ is the area moment of inertia of the plate; $h$ is the thickness of the plate and $l$ is the length of the plate. We vary $m$ across orders of magnitude, $m\in[10^{-4},10^2]$, for three fixed values of $k$. Relatively few values of $k$ are chosen because for this FSI system, this stiffness parameter sets the equilibrium balance between the mean flow stresses on the plate and the mean deflection profile. The specific values of $k$ and their justification are discussed in section \ref{sec:results} below, but broadly they are selected to yield modest mean deflection values while still allowing non-trivial fluid-structure interaction dynamics to arise. The mass value $m$ is then varied to systematically tune the intrinsic structural timescales (natural frequencies) relative to the underlying vortex-shedding frequency (for the reference rigid case), to quantify the ensuing fluid-structure interplay.

\subsection{Numerical scheme}
We use the strongly coupled immersed boundary projection algorithm for deforming surfaces formulated by Goza and Colonius \cite{goza2017} and modified for linear Euler-Bernoulli beams by Balasubramanian and Goza \cite{balasubramanian2022}. The governing equations being solved are the Navier-Stokes, continuity, Euler-Bernoulli, and the no-slip constraint equations, written in dimensionless form as
\begin{gather}
\frac{\partial \textbf{u}}{\partial t}+\textbf{u}\cdot\nabla \textbf{u}=-\nabla p +\frac{1}{Re}{\nabla}^2\textbf{u}+\int_{\Gamma}\textbf{f}(\boldsymbol{\chi}(s',t))\delta(\boldsymbol{\chi}(s',t)-\textbf{x})ds',
\label{eqn:NS} \\
\nabla\cdot\textbf{u}=0, \label{eqn:continuity} \\
\frac{\partial^2}{\partial{s'}^2}\left(k({s'})\frac {\partial^2 {\boldsymbol{\chi}}}{\partial{s'}^2  }\right)=-m({s'})\frac{\partial^2{\boldsymbol{\chi}}}{\partial {t}^2} + {\textbf{f}({\boldsymbol{\chi}},{t})} + {\textbf{g}({\boldsymbol{\chi}})}, \label{eqn:EB} \\
\int_{\Omega}\textbf{u}(\textbf{x})\delta(\textbf{x}-\boldsymbol{\chi}(s',t))d\textbf{x}=\frac{\partial \boldsymbol{\chi}(s',t)}{\partial t}. \label{eqn:no-slip}
\end{gather}
In the above, $\Gamma$ denotes the immersed flat plate and the fluid domain is indicated by $\Omega$. The displacement of the flat plate (parametrized by a variable $s'$) is given by the Lagrangian coordinate $\bm\chi(s,t)$, and $\textbf{x}$ indicates the Eulerian flow domain coordinates. The quantities $\textbf{u}$ and $\textbf{p}$ denote the flow velocity and pressure, respectively. The structural parameters $k,$ $m$ are the dimensionless stiffness and mass ratios defined in (\ref{eqn:params}), and $\textbf{f}(\bm\chi,t)$ is the surface stress acting on the plate. The variable $g(\chi)$ is a body force that acts on the plate (e.g., due to gravity). In the context of this article, this body force term will be used to apply a pre-stress to the structure, as described in section \ref{sec:base-state-compute}.

We recast equations (\ref{eqn:NS})--(\ref{eqn:continuity})  in a vorticity-streamfunction formulation as in reference \cite{taira2008} and spatially discretize the resulting equations using a standard second-order finite volume scheme. The spatially discrete equations are given as (see \citet{goza2017} for more details)
\begin{gather}
C^{T}C\dot{s}=-C^{T}N(Cs)+\frac{1}{Re}C^{T}LCs-C^{T}E^{T}(\chi)f,
\label{eqn:NSdiscrete} \\
M\dot{\zeta}=-K\chi+Qg+QW(\chi)f,
\label{eqn:EBdiscrete} \\
\dot{\chi}=\zeta,
\label{eqn:EB1discreet} \\
 0=E(\chi)Cs-\zeta.                    
\label{eqn:no-slipdiscrete}
\end{gather}
In equations (\ref{eqn:NSdiscrete}-\ref{eqn:no-slipdiscrete}), $C$ is the spatially discrete curl operator and $s$ is the streamfunction. $N(Cs)$ is the spatially discrete advection operator and $L$ is the discrete approximation of the Laplacian operator. The matrices $E,$ $E^{T}$ are the discrete versions of the regularization and smearing operators employed in the immersed boundary formulation. For the structural equation (\ref{eqn:EBdiscrete}), $K\chi$ is the internal stress on the beam, $Qg$ is the externally imposed body force on the beam and $QW(\chi)f$ is the force on the beam due to the surrounding fluid. The structural state variables $\chi$ and $\zeta$ are the spatially discrete versions of beam displacement and velocity.

Dirichlet boundary conditions are utilized for the streamfunction, facilitating linear solves with fast Fourier transforms. This boundary treatment is enabled by a multi-domain approach described in \citet{taira2008}. For time marching the flow equations, an Adams-Bashforth time-discretization scheme is used for the nonlinear term and the diffusive term is treated implicitly in time using a trapezoidal scheme. The Euler-Bernoulli equation (\ref{eqn:EB}) is spatially discretized using a second-order finite element method and time-discretized using the implicit (zero-dissipation) Newmark scheme. The no-slip constraint (\ref{eqn:no-slip}) is enforced implicitly at each time step, yielding a nonlinear algebraic set of equations solved iteratively via a Newton method, where each solve of the linearized system at a given iteration is performed by an  efficient block-LU algorithm \cite{goza2017}. 

\subsection{Base-state initialization} \label{sec:base-state-compute}
 The base-state for the FSI system of equations (\ref{eqn:NSdiscrete})--(\ref{eqn:no-slipdiscrete}) corresponds to the steady state solution with all time derivatives set to zero. In other words, if $r(y)$ corresponds to the right hand side of equations (\ref{eqn:NSdiscrete})--(\ref{eqn:no-slipdiscrete}) with $y=[s,\chi,\zeta,f]^T$, then $y_b$ (base-state) is such that $r(y_b)=0$. Our aim in this work to focus on the dynamic interplay between the flow and structure, rather than the effect of the mean camber on the flow. As such, we introduce a pre-stress into the structural equations of motion, via the body force term $Qg$, such that the base state of the fully coupled system is associated with a zero deflection state. That is, the equilibrium of the rigid-plate system equations (\ref{eqn:NSdiscrete}), (\ref{eqn:no-slipdiscrete}) is computed first to provide a base streamfunction and surface stress, $s_b$ and $f_b$, respectively. These base state quantities of the rigid-body system are then the base quantities for the fully coupled system (with $\chi_b$, $\zeta_b\equiv0$) provided that the body force is set to $Qg = -QWf_b$. We emphasize that this formulation yields a formal base state of the fully coupled system, and one that allows for a direct exploration of the dynamic evolution of the FSI system starting from an undeflected state.  
 
 Note that even for the rigid-body system, the base flow state at this Reynolds number and angle of attack is unstable. As such, we use the fixed point algorithm that leverages the nonlinear time-stepping software described in \citet{ahuja_rowley_2010,kelley,tuckerman} to compute the base-state.  A modified Newton algorithm is used to advance the guess for $y_b$ with the Newton update computed as products between the Jacobian matrix and the initial condition for the latest guess of the base state. The Jacobian-vector products are approximated using finite differences and computed in a matrix-free way using GMRES. At the end of this base-state computation we obtain $s_b,f_b$ from the rigid-body system, which yields the undeformed equilibrium of the fully coupled FSI system under the pre-stress $Qg=-QWf_b$.

These base-state computations provide the initial condition for our nonlinear simulations. All simulations are perturbed from this base state using a small body force in the flow for initial time, and the time evolution of the FSI state is computed from the high-fidelity solver.

\section{Results}\label{sec:results}
\subsection{Baseline flow configuration} \label{sec:baseline_results}
\begin{figure}[h]
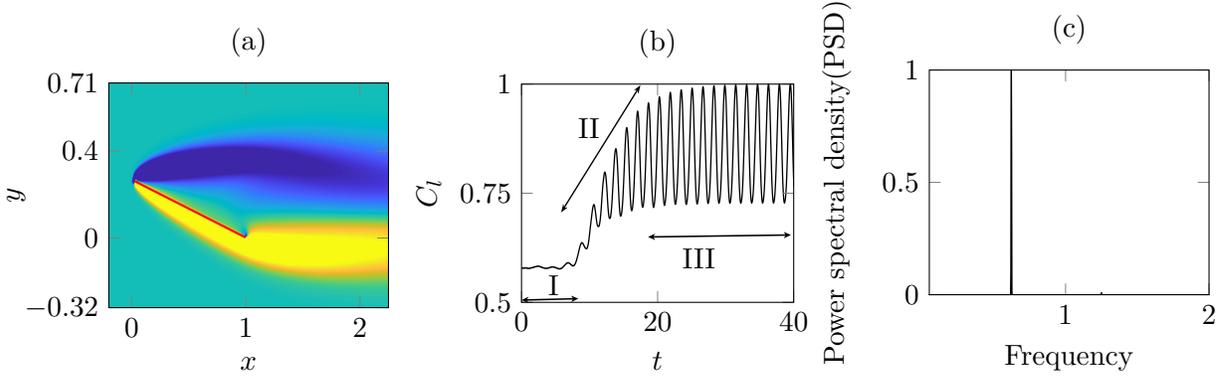

\begin{subfigure}[b]{0.33\textwidth}
\input{figures/overviewfigures_1/base.tex}
\label{fig:fig2a}
\end{subfigure}
\begin{subfigure}[b]{0.32\textwidth}
\input{figures/overviewfigures_1/baseregimes.tex}
\label{fig:fig2b}
\end{subfigure}
\begin{subfigure}[b]{0.31\textwidth}
\input{figures/overviewfigures_1/fft}
\label{fig:fig2c}
\end{subfigure}

\caption{(a): Vorticity contour of the unstable base state at $Re=500$, angle of attack $15^{\circ}$, (b): Time trace plot of the lift coefficient $C_l$ associated with the rigid-baseline case. The early time phase ($I$), saturation phase ($II$) and limit cycle phase ($III$) are shown, (c): Normalized power spectral density associated with the $C_l$ time trace}
\label{fig:baseline}
\end{figure}

Figure \ref{fig:baseline}(a) shows the vorticity contour associated with the base state for flow past a rigid flat plate at the angle of attack =$15^{\circ}$ and $Re=500$. The leading- and the trailing-edge vortex wakes are slightly asymmetric with respect to the centerline owing to the large angle of attack associated with the plate. The base state is unstable: any perturbation at $t=0$ leads to oscillations that depart the base state and settle into limit-cycle oscillations. 

Figure \ref{fig:baseline}(b) shows the lift coefficient ($C_l$) associated with the entire plate, with respect to time. At $t=0$, the system is perturbed via a small-amplitude body force and this sets up an evolution in time for the instability growth. The asymmetric wake develops small oscillations in the linear regime I (see figure \ref{fig:baseline}(b)). The unstable oscillations grow in amplitude and subsequently advect downstream. By the end of phase I, the growing oscillations in the vortices have reached a magnitude sufficient enough to change the $C_l$ on the plate, leading to a departure from the base lift coefficient of $C_l\approx0.58$. 

In phase II, the vortex shedding is associated with larger-amplitude oscillations, with a corresponding increase in the mean lift. The vorticity distribution at the base state ($t=0$) is modified by the increasing magnitude of vortex oscillations and under the action of freestream advection. The vortices near the leading-edge (LEV) and the trailing-edge (TEV) are swept downstream. Throughout the saturation phase (II), the vortex-oscillations and freestream advection continue until a dynamic equilibrium is achieved between the two processes. By the end of II, the vortex and lift oscillations asymptote to a periodic limit cycle ($t\approx20$ in figure \ref{fig:baseline}(b)), marking the start of Phase III. In this phase, the leading-edge vortex (LEV) and the trailing-edge vortex (TEV) are shed alternately in time. The frequency associated with the shedding shows a gradual increase from I to II, before settling to a constant value in III. The frequency associated with the shedding in III is computed through a power spectral density of the $C_l$ time trace, sampling the portions of the signal once the limit cycle is reached (see figure \ref{fig:baseline}(c)). The frequency shown in figure \ref{fig:baseline}(c) will be referred to as the baseline shedding frequency ($f_s$), and is evident from the power spectral density plot as the dominant frequency $f_s=0.623$. 

\begin{figure}
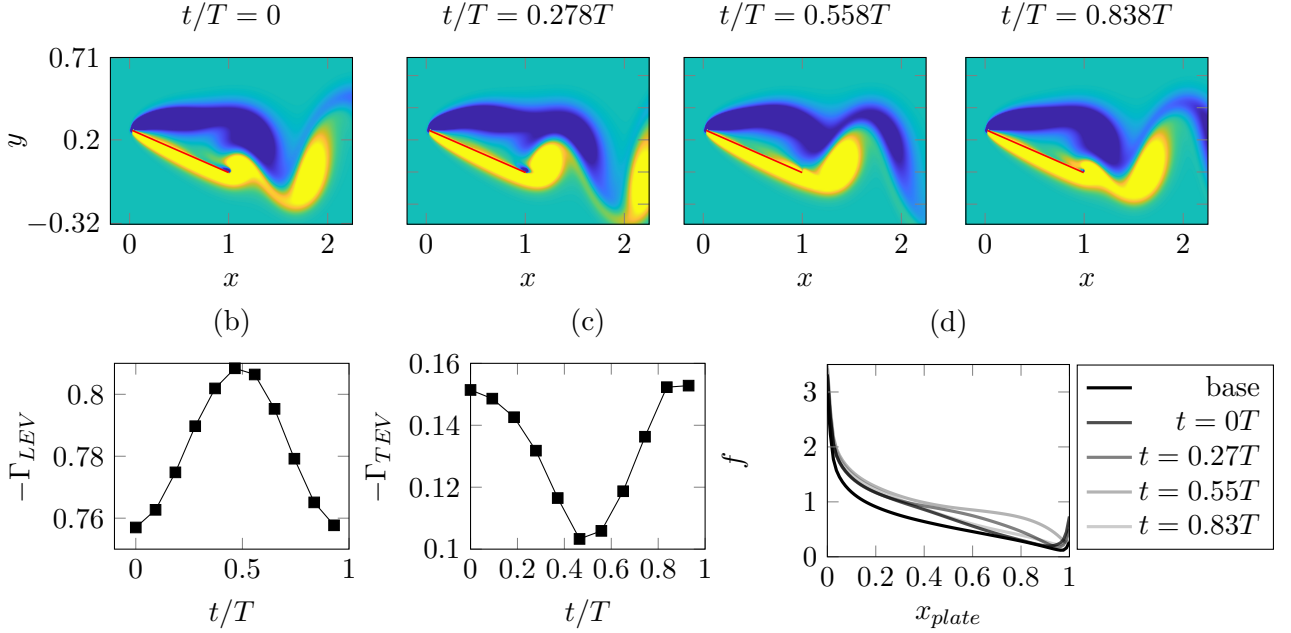


\begin{subfigure}[b]{0.31\textwidth}
\input{figures/overviewfigures_1/rig1.tex}
\end{subfigure}
\begin{subfigure}[b]{0.22\textwidth}
\input{figures/overviewfigures_1/rig2.tex}
\end{subfigure}
\begin{subfigure}[b]{0.22\textwidth}
\input{figures/overviewfigures_1/rig3.tex}
\end{subfigure}
\begin{subfigure}[b]{0.22\textwidth}
\input{figures/overviewfigures_1/rig4.tex}
\end{subfigure}

\begin{subfigure}[b]{0.28\textwidth}
\input{figures/overviewfigures_1/levrigid.tex}
\end{subfigure}
\begin{subfigure}[b]{0.28\textwidth}
\input{figures/overviewfigures_1/tevrigid.tex}
\end{subfigure}
\begin{subfigure}[b]{0.4\textwidth}
\input{figures/overviewfigures_1/stress.tex}
\end{subfigure}

\caption{Top row: Vorticity snapshots of flow past a rigid plate at the equally spaced time instances $t/T=0, 0.278, 0.55, 0.83$ in a vortex-shedding cycle,
(b,c): Variation of  circulation strength associated with the leading-edge vortex (b) and the trailing-edge vortex (c) over a vortex-shedding cycle,
(d): Resultant surface stress ($f$) distribution along the length of the plate, at the base state (dark line) and at time instances ($t/T=0, 0.27, 0.55, 0.83$ shown by decreasing opacity) over a baseline vortex-shedding cycle, }
\label{fig:baselinecycle}
\end{figure}

Figure \ref{fig:baselinecycle} (top row) shows the vorticity contours sampled at specific time instances within the rigid-baseline lift cycle , where $T$ is the time period of one limit cycle oscillation. The circulation strengths associated with the clockwise leading-edge vortex (LEV) and the anti-clockwise trailing-edge vortex (TEV) strengths are shown in figures \ref{fig:baselinecycle}(b), (c). Figure \ref{fig:baselinecycle}(d) also shows the resultant of surface stress that acts along the length of the plate. That is, denoting $\textbf{f}=[f_x, f_y]^T$ as the surface stress vector along the plate, then $f:=||\textbf{f}||$ is plotted along the length of the plate in figure \ref{fig:baselinecycle}(d) at various instances within a vortex-shedding cycle. The domain and procedure for the circulation strength computations are outlined in Appendix A. 

At $t=0T$, the circulation  associated with the TEV is at its maximum strength within the cycle, while the LEV is at its minimum strength (see figure \ref{fig:baselinecycle}(b),(c)). It is visually evident from the vorticity contour (at $t/T=0$) that the TEV has just formed to its maximum size and is beginning to advect downstream. In contrast, the LEV above the plate has just advected and is beginning its roll-up process. At $t/T=0$, figure \ref{fig:baselinecycle}(d) shows that the magnitude of resultant surface stress distribution along the length of the plate, on average, is slightly higher compared to the distribution seen at the base state.

From $t=0T$ to $t=0.278T$, the circulation strength associated with the TEV decreases while there is an increase in the circulation strength associated with the LEV. The full-strength TEV formed at $t/T=0$ has advected away from the plate to a position downstream, while the LEV starts its roll-up process (visually evident from the vorticity snapshot at $t/T=0.278$ from figure \ref{fig:baselinecycle}). The LEV roll-up increases its curvature/strength above the plate leading to an more pronounced low-pressure region immediately above the plate. Figure \ref{fig:baselinecycle}(d) shows that an increased magnitude of surface stress acts on the plate at this instance. From $t=0.27T$ to $t=0.558T$ the LEV continues its formation phase (as evident from figure \ref{fig:baselinecycle}(b)) and increasingly occupies more area on the suction side of the plate. The increasing LEV strength is again associated with an increased magnitude of surface stress on the plate (as evident from the profile for $t=0.558T$ in figure \ref{fig:baselinecycle}(d)). At $t=0.558T$, the positive circulation above the plate is at its maximum within the cycle, indicating maximum roll-up of the LEV. At the same instance, the negative circulation associated with the TEV is at its minimum (see figure \ref{fig:baselinecycle}(c)) within the cycle. This low circulation strength is associated with the TEV fully advecting away from the plate by $t=0.558T$.

From $t=0.558T$ to $0.838T$, the circulation strength associated with the TEV increases while there is a drop in the LEV circulation. Visually, a new TEV is seen to begin its formation process while the fully formed LEV advects downstream. The counter-clockwise circulation associated with the TEV formation helps in advecting the LEV, leading to a reduced clockwise circulation above the plate. There exists an increased region of reverse flow above the plate during this process, inevitably leading to a reduced magnitude of surface stress throughout the length of the plate (though still on average greater than the base state). The vortex-shedding processes from $t=0.558T-0.838T$ are seen to be detrimental to lift. The lift coefficient continues to drop beyond this time from $t=0.838T$ to $t=T$ until a new LEV begins its formation cycle.

Using this baseline vortex-shedding process and its associated lift behavior, the following sections probe how the evolution from the formally computed base state changes, both for early time and in the long-time limit, when the plate is made flexible. This study will be addressed with plate parameters that are systematically varied so that structural natural frequencies are tuned relative to the identified $f_s$ reference vortex-shedding frequency.

\subsection{ Parametric space}

\begin{figure}
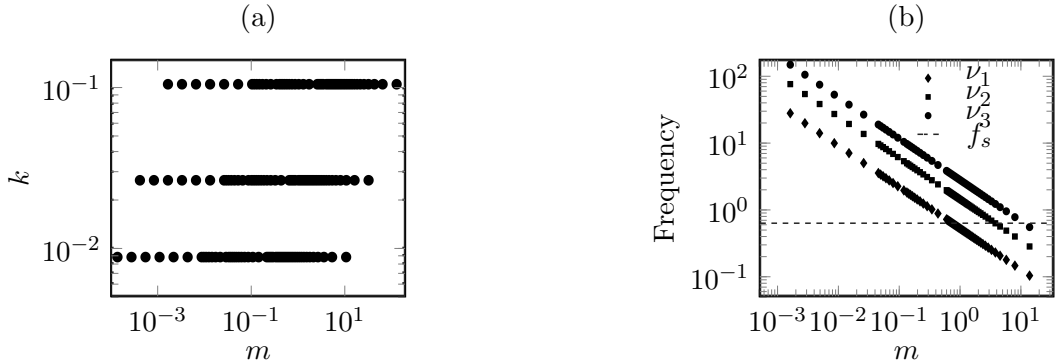


\begin{subfigure}[b]{0.47\textwidth}
\input{figures/overviewfigures_1/mkspace.tex}
\end{subfigure}
\begin{subfigure}[b]{0.47\textwidth}
\input{figures/overviewfigures_1/vacuumfreq.tex}
\end{subfigure}

\caption{(a): $k$-$m$ parameters of the flexible plate for which numerical simulations are run, (b): Fundamental frequencies of the in-vacuo Euler-Bernoulli beam for the range of $m$ considered, dashed horizontal line indicates the baseline vortex-shedding frequency}
\label{fig:parametricspace}

\end{figure}

 For the  FSI simulations of flow past a flexible flat plate, we consider the $k$-$m$ parametric space as in figure \ref{fig:parametricspace}(a). The black markers denote the specific parameter combinations $(m,k)$ that are simulated. The selections include three values of $k$, and a range of $m$ for each $k$ so that the plate natural frequency is systematically varied with respect to the fundamental vortex-shedding frequency of the baseline (rigid) case, $f_s$.
 
 For a fixed mass $m$, the stiffness $k$ dictates the steady state static response of the structure to the mean lift force acting on the plate, in its aerodynamic configuration. (In other words, the mean lift force dictates the time-averaged camber shape of the flexible structure). Note that while the dynamics can play a role in altering the mean flow state,  the results below demonstrate that a fixed $k$ yields a roughly constant mean deformation profile. As such, this stiffness value $k$ is used as a proxy for fixing the mean structural deformation, to focus instead on the flow changes induced by structural dynamics. The values of $k$ are chosen such that the induced mean camber is not too large so as to invalidate the linear Bernoulli model used (this constraint sets up the lower limit on $k$ for this study). The upper limit on $k$ is dictated by our search for parameters that enable structural oscillations to have a significant impact on the flow processes compared with the rigid case. Higher $k$ values predominantly show a one-way response: the structure is excited by the shedding processes, but the triggered small-amplitude structural excitations in turn only  minimally affect the flow. Keeping these preliminary observations in mind, we restrict our parametric space for three different values of $k$ as shown in figure \ref{fig:parametricspace}(a). 
 
As the primary focus of the study is to explore the coupled dynamics of the system about a fixed mean camber (attempting to isolate the role of relative frequency alignment from static stiffness-induced effects), the  majority of the study will focus on a single $k$ value, $k=0.026$. This value was chosen because it yielded a modest mean deflection value ($\approx 0.02$) while providing distinct FSI from the rigid reference case. Results at the other $k$ values are provided in Appendix \ref{sec:App} to demonstrate  qualitatively compatible trends in varying the mass. 

For a fixed $k$, we consider a range of $m$ such that the vacuum fundamental frequencies of the plate are aligned differently with respect to the baseline shedding frequency $f_s$. This change in natural frequency is illustrated in figure \ref{fig:parametricspace}(b), where we show the first three fundamental frequencies of the in-vacuo Euler-Bernoulli beam for the range of considered $m$ (the dashed line represents the baseline vortex-shedding frequency $f_s$ and is provided for reference). Of course, the in-vacuo scaling of the natural frequencies may not accurate reflect the timescales of vibration of the full FSI system; an assessment of this outcome will be made in the results sections below. 

In the rest of the manuscript, we address the following sets of questions for the FSI system. First, what is the early-time evolution of the FSI system from the formal base state, and is it associated with a coupled fluid-structure or flow-excited instability? Does this outcome change for different structural parameters? Second, for the $m$ parameters chosen, can we identify distinct regimes associated with the coupled dynamics in the long-time limit?  What are qualitative changes to the observed structural mode and vortex-shedding behavior for any identified regimes? What is the relationship between intrinsic structural and baseline vortex-shedding timescales that are associated with these various regime changes? Third, how do the structural dynamics affect the vortex-shedding behavior (and associated circulation properties) from what occurs for the reference rigid case detailed in section \ref{sec:baseline_results}? How does this interplay conspire to alter the lift behavior of the plate? 

\subsection{Overview of FSI dynamics and behavioral regimes}
\subsubsection{Early time dynamics}
\label{sec:early_time_overview}
We first investigate the early-time behavior of the dynamics at small times close to the base-state of the configuration (time region I in figure \ref{fig:baseline}(b)). Inspecting this time window allows us to isolate the linear, small-amplitude instabilities that drive a departure from the base state before nonlinear effects arise.

We first consider the aerodynamic lift coefficient ($C_l$) acting on the plate and the midpoint displacement of the flat plate ($\chi_{mid}$). The window I (see figure \ref{fig:baseline}(b)) is sampled until the time instance of departure from the base-state (this is approximately the instance when the $\overline{C_l}$  exceeds ${C_l}_{base}$ for the first time) and the duration of the interval varies for every $m$ value we consider. Over this time window, we compute the wavelet spectrogram of the signal and extract the frequency corresponding to the maximum power shown in the spectrogram. For each $m$, the obtained value of the dominant frequency is plotted in figure \ref{fig:mean-overview}(a) for both the $C_l$ as well as the $\chi_{mid}$ signals. We note that since the linear-instability portion of the signal only consists of a few cycles, the spectrogram peaks are relatively broad and we indicate the width of the spectral peak (within one standard deviation of the peak magnitude) using the vertical bars at each marker.

For both $C_l$ and $\chi_{mid}$, the spectral content is nearly constant across the wide range of $m$ values, chosen to align the reference rigid vortex-shedding frequency with three distinct structural modes. The peak frequency in both $\chi_{mid}$ and $C_l$ is within $[0.45, 0.53]$. This range of values straddles the value observed for $C_l$ in the rigid case, indicated by the dashed line. (Note that this frequency of the onset of vortex shedding for the rigid case is lower than the frequency observed after nonlinear saturation occurs and the limit cycle is reached, $f_s=0.623$). The commonality of this frequency value for early times across a wide range of mass values $m$, which will be shown to yield notably distinct dynamics over long times, and the closeness of this common frequency value to the reference rigid case, suggests a common linear instability mechanism in all cases. That is, even for the cases with a compliant structure, the data suggests that it is the vortex-shedding instability triggered by bluffness of the structure to the flow that initiates oscillations of the FSI system. Even in the rigid-body setting, these oscillations go on to induce nonlinear modulation effects. Nonlinearity for the rigid case yields a different mean lift than the base state lift as well as unsteady periodic vortex-shedding processes. In the remaining sections, we will probe how nonlinearities manifest for the FSI system as a function of the structural mass $m$.

\subsubsection{Long-time dynamics}
\label{sec:long_time_overview}

\begin{figure}
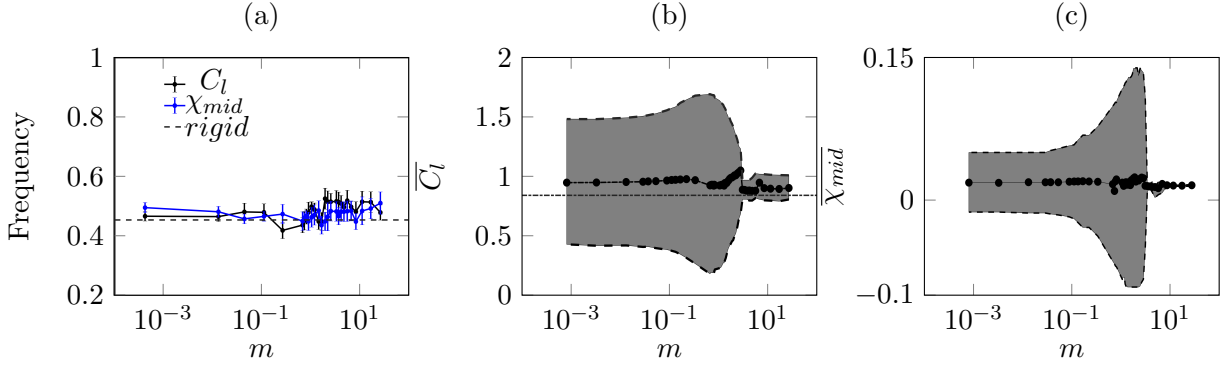

\begin{subfigure}[b]{0.32\textwidth}
\input{figures/overviewfigures_1/earlytime.tex}
\end{subfigure}
\begin{subfigure}[b]{0.32\textwidth}
\input{figures/overviewfigures/Fig5b.tex}
\end{subfigure}
\begin{subfigure}[b]{0.32\textwidth}
\input{figures/overviewfigures/Fig5c.tex}
\end{subfigure}
\caption{(a): Frequency associated with the early-time portion of the $Cl$ and $\chi_{mid}$ signals (see main text for details). (b,c): Time-averaged $C_l$ (b) and $\chi_{mid}$ (c) 
(black markers) versus $m$. The window of time averaging starts once the long-time attractor is reached  (see figure \ref{fig:baseline} for the relevant time instances).
In (b), (c), the dashed lines on either side of the black markers denote the extrema 
over the time window used to compute the average. The dot-dashed line at $C_l=0.8431$ represents the mean $C_l$ value for the rigid-baseline case.}

\label{fig:mean-overview}
\end{figure}

We consider in figure \ref{fig:mean-overview}(b)--(c) the long-time behavior of the lift and midpoint displacement of the flow-plate system. All the $\overline{C_l}$ markers lie above the rigid reference mean lift (and, by extension, the base state lift). The magnitude of the peak oscillations (relative to the mean) is large for the cases $m<2.76$ and becomes almost negligible for $m>2.76$. There are slight variations in $\overline{C_l}$ and $C_l$ extrema, that exist across the black markers. The far left of the parametric space (low $m$) has a nearly constant mean lift irrespective of $m$ until  about $m \approx 0.01$. There is a slight increase in $\overline{C_l}$ as we reach $m \approx 0.2$ and a slight drop in the mean lift from $m \in [0.2,0.6]$. The mean lift on the plate is at its highest at $m=2.76$, with $\overline{C_l}=1.05$. As we consider $m>2.76$, there is a sharp drop in the mean lift and a nearly constant reduced mean lift value of $\overline{C_l}\approx0.9$ is observed for $m>2.76$. 

For the lift extrema, both the  dashed lines (lift maximum on top and lift minimum on the bottom) are nearly constant at  low $m$ ($m<0.1$). The peak-to-peak amplitude in the extrema is large over these small mass values, on the order of the mean lift value itself. With increasing $m$,  $m\in[0.02,0.66]$, there is an increase in the peak-to-peak $C_l$ excursions. The 
peak-to-peak amplitude in the lift extrema is maximal 
at $m=0.66$ (whereas the maximum in the mean lift coefficient, $\overline{C_l}$, occurs at $m=2.76$). The peak-to-peak $C_l$ amplitude decreases over $m\in[0.66,2.76]$. For $m>3$, the magnitude of $C_l$  peak-to-peak excursions are considerably smaller than for low $m$

For the midpoint displacement $\chi_{mid}$, the black markers in figure \ref{fig:mean-overview}(c) denote a roughly constant $\overline{\chi_{mid}} \approx 0.02$ for all $m$ values considered. (This outcome is expected, and was indeed designed into the problem setup by fixing the stiffness $k$ across this set of cases). However, the peak oscillation amplitude (dashed lines) vary across $m$. The oscillation magnitude is nearly constant for low mass values, $m<0.1$. Over $m\in[0.1,2.76]$, the top and the bottom dashed lines diverge before reaching  maximum/minimum values  at $m=2.76$. For $m=2.76$, the plate exhibits oscillations of the largest magnitude (as was observed for the lift extrema). For $m>2.76$, we see that the structural oscillations about the mean are small in amplitude (again, similar to what was observed in the $C_l$ extrema).

The quantities $\overline{\chi_{mid}}$, peak-to-peak $\chi_{mid}$ extrema and $\overline{C_l}$ have their maximal value at $m=2.76$, whereas the the peak-to-peak $C_l$ extrema  show their respective maximum/minimum at $m=0.66$. The maximum structural oscillations are therefore associated with maximum mean lift but not with maximum magnitude $C_l$ oscillations. The mechanistic reasoning behind this discrepancy will be explored further, later in the manuscript.

\begin{figure}
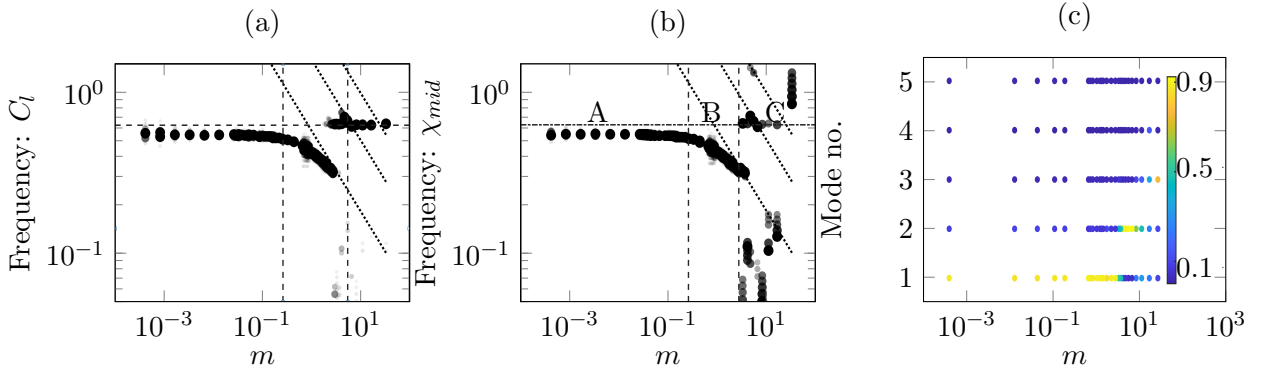

\begin{subfigure}[b]{0.32\textwidth}
\input{figures/overviewfigures_1/spectralcl.tex}
\end{subfigure}
\begin{subfigure}[b]{0.32\textwidth}
\input{figures/overviewfigures_1/spectralmidpt.tex}
\end{subfigure}
\begin{subfigure}[b]{0.32\textwidth}
\input{figures/overviewfigures_1/projection.tex}
\end{subfigure}
\caption{Information about the power spectral density (PSD) associated with the long-time $C_l$ (a) and $\chi_{mid}$ (b) oscillations (see main text for details). The vertical dashed lines demarcate the distinct regimes of behavior, described in the main text. The horizontal dashed/dotted line marks the frequency for the rigid baseline case, $f_s=0.623$. In (a), (b) the dashed lines are the first, second and third fundamental frequencies of the Euler-Bernoulli beam in vacuum,
(c): Projection coefficient (indicated by color shade) associated with long-time limit cycle structural dynamics (see main text for details).}
\label{fig:spectral-projection}
\end{figure}

The dynamics can be probed beyond an assessment of the mean and maximal excursion from this mean for $C_l$, $\chi_{mid}$, by quantifying the spectral content in $C_l$, $\chi_{mid}$ as well as the structural modes observed throughout each simulation. 
To this end, figures \ref{fig:spectral-projection}(a), \ref{fig:spectral-projection}(b) show frequency information associated with the long-time $C_l$ and $\chi_{mid}$ oscillations, respectively. At each $m$ value, results from a power spectral density (PSD) analysis are plotted as markers, with the largest, darkest markers indicating the frequency values of largest energy and smaller, lighter markers to frequency values of lower energy. (Thus, there are generally multiple markers for each $m$ value). The marker size and opacity is set relative to the largest energy at each $m$ value, and is scaled from the most to least energetic frequency at each $m$ value. Only the frequencies corresponding to the six most energetic peaks in the PSD are shown as non-white markers. The time window for the PSD computations start once the long-time attractor is reached. 

For low mass values of $m<0.26$, the dominant frequency associated with $C_l$, $\chi_{mid}$ dynamics is nearly constant irrespective of $m$. The spread in opacity and size is small, and the dominant frequency value is slightly below the rigid-baseline frequency $f_s$. For $m\in[0.26,2.76]$, the dominant frequency shows a decreasing trend (roughly scaling as the square root of $m$, with the spread about this peak small for both $\chi_{mid}$ and $C_l$). The dominant frequencies are seen to increasingly align closely with the first in-vacuo fundamental frequency (leftmost dotted lines in figure \ref{fig:spectral-projection}(a), (b)), especially near $m=2.76$. For $m>2.76$, the dominant frequency associated with the $C_l$ dynamics abruptly shifts from the in-vacuo fundamental frequency to the baseline vortex-shedding frequency, $f_s$. A similar shift occurs in the $\chi_{mid}$ dynamics for $m>2.76$, albeit with a larger spread in the frequency around $f_s$. The large spread in the frequencies for $\chi_{mid}$ possibly indicate the presence of structural harmonics triggered by flow forcing, which in turn have a small impact on $C_l$ (no signatures of this range of frequencies is visible on the $C_l$ plot).

In figure \ref{fig:spectral-projection}(c), we show the projection coefficient of the  long-time plate dynamics onto the various Euler-Bernoulli mode shapes. For every mass $m$, the instantaneous plate displacement field about the mean camber is sampled at equal intervals in time and projected onto the in-vacuo eigenvectors of the beam. The instantaneous projection coefficients are then averaged across the sampled time window. (To facilitate a clearer comparison across each mass value $m$, the averaged projection coefficients are  scaled by the maximum value of the coefficient obtained for each $m$ over the sampling time window). The figure shows that mode one is dominant for $m\in[0.0004,2.76]$. For $m\in[2.76,4.14]$, there is a gradual fade in mode one presence and at $m=7$, mode one is entirely inactive. Mode two is most prominent at $m=6.626$. For $m>6.626$, mode two presence begins to fade with $m=26.16$ showing oscillations dominated by mode three.

 From the noted trends in long-time flow-structure quantities, we categorize the dynamics over the considered mass values into three qualitatively distinct 
 regimes (A, B and C). These regimes are indicated in figure \ref{fig:spectral-projection}(a)--(b), and summarized in what follows using the results from figures \ref{fig:mean-overview}--\ref{fig:spectral-projection}. For low values of $m\in[0.0004,0.26]$ (A), there is a near-constancy in the mean and frequency of $C_l$ and $\chi_{mid}$. The dynamics are essentially independent of $m$ within this regime, and the structural behavior is predominantly mode one in nature. For $m\in[0.26,2.76]$ (B), the peak-to-peak amplitude in the maximal excursions from the mean for $\chi_{mid}$ and $C_l$ vary across this mass range. Moreover, the mean lift-coefficient ${C_l}$ generally increases over this regime B with increasing $m$. The frequency associated with the dominant dynamics decreases from $m=0.26$ to $m=2.76$, with the values increasingly becoming aligned with the first structural in-vacuo fundamental frequency. Within this regime, the peak-to-peak amplitude of $C_l$ excursions show an increasing then decreasing trend with $m$. The maximal mean lift $\overline{C_l}$ and maximal  excursions from the mean structural displacement occur once the lift oscillation magnitude decreases. This discrepancy suggests a complicated interplay between mechanisms that drive lift oscillations and those that drive structural oscillations and mean lift benefit. For $m>2.76$ (C), the excited structural modes transition from the first mode to higher modes (mode two at $m=5.52$ and mode three at $m=26.16$). The higher mode structural oscillations are small in magnitude and have a negligible impact on $C_l$ relative to the rigid baseline case.

\subsubsection{Qualitative summary of behavioral regimes, questions for the rest of the article}
\label{sec:overview_summary}

\begin{figure}
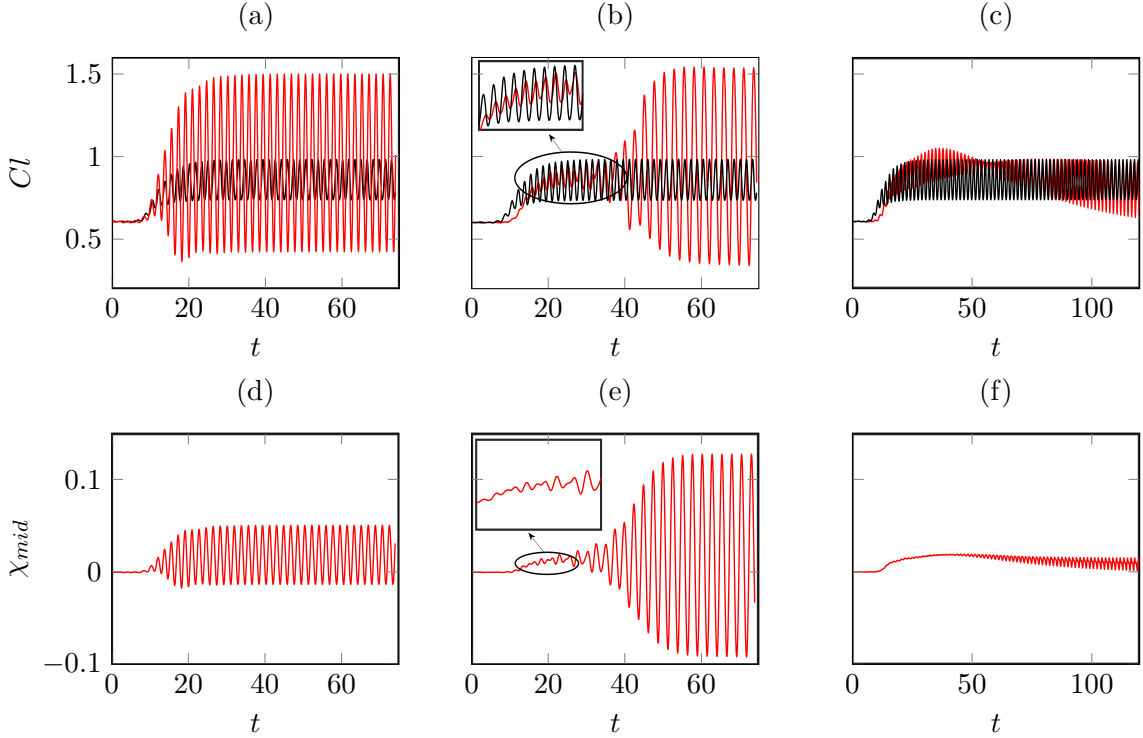

\begin{subfigure}[b]{0.34\textwidth}
\input{figures/overviewfigures_1/amasscltime.tex}
\end{subfigure}
\begin{subfigure}[b]{0.3\textwidth}
\input{figures/overviewfigures_1/mode1cltime.tex}
\end{subfigure}
\begin{subfigure}[b]{0.3\textwidth}
\input{figures/overviewfigures_1/mode2cltime.tex}
\end{subfigure}
\begin{subfigure}[b]{0.34\textwidth}
\input{figures/overviewfigures_1/amassmidpttime.tex}
\end{subfigure}
\begin{subfigure}[b]{0.3\textwidth}
\input{figures/overviewfigures_1/mode1midpttime.tex}
\end{subfigure}
\begin{subfigure}[b]{0.3\textwidth}
\input{figures/overviewfigures_1/mode2midpttime.tex}
\end{subfigure}

\caption{Time trace plots of the $C_l$ (a-c) and $\chi_{mid}$ (d-f) signals. Each column corresponds to a different mass value $m$, as:  $m=0.0132$ (a,d), $m=1.44$ (b,e), and $m=5.52$ (c,f). 
}
\label{fig:time-evolution}
\end{figure}

\begin{figure}
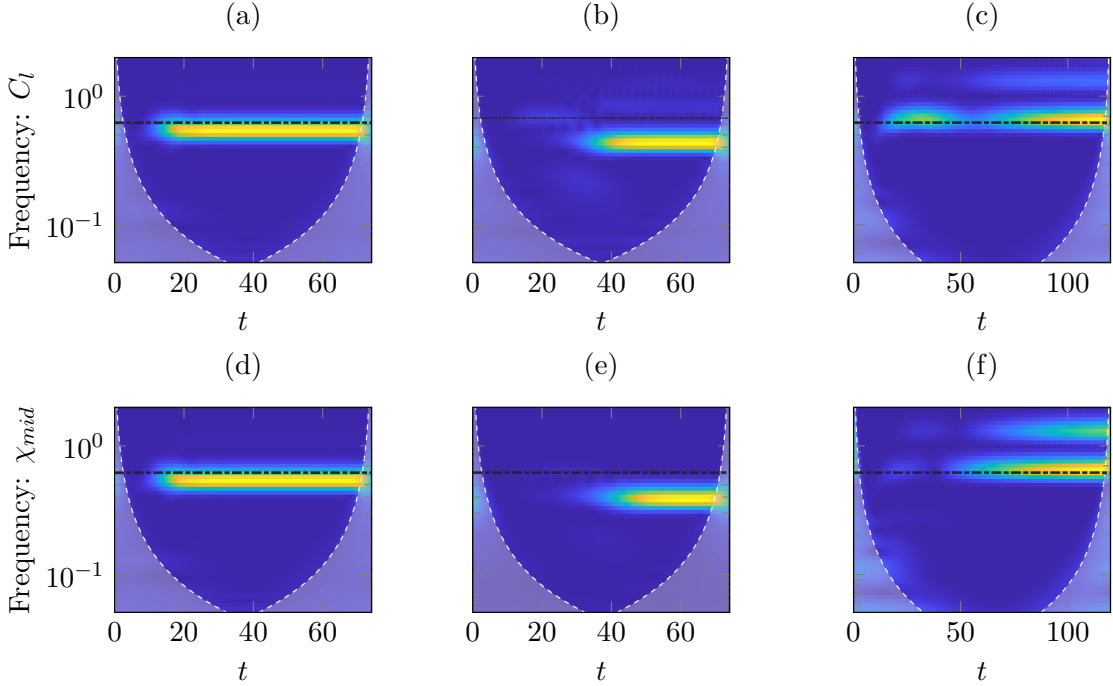

\begin{subfigure}[b]{0.34\textwidth}
\input{figures/overviewfigures_1/amassclspec.tex}
\end{subfigure}
\begin{subfigure}[b]{0.3\textwidth}
\input{figures/overviewfigures_1/mode1clspec.tex}
\end{subfigure}
\begin{subfigure}[b]{0.3\textwidth}
\input{figures/overviewfigures_1/mode2clspec.tex}
\end{subfigure}
\begin{subfigure}[b]{0.34\textwidth}
\input{figures/overviewfigures_1/amassmidptspec.tex}
\end{subfigure}
\begin{subfigure}[b]{0.3\textwidth}
\input{figures/overviewfigures_1/mode1midptspec.tex}
\end{subfigure}
\begin{subfigure}[b]{0.3\textwidth}
\input{figures/overviewfigures_1/mode2midptspec.tex}
\end{subfigure}

\caption{ Wavelet spectrograms ($C_l$: (a-c), $\chi_{mid}$: (d-f)). Each column corresponds
to a different mass value $m$, as: $m = 0.0132$ (a,d), $m = 1.44$ (b,e), and $m = 5.52$ (c,f). The dashed line is the peak frequency from a power spectral density analysis of the lift in the rigid reference case. The legends for the colorbar is omitted for cleanness.}
\label{fig:time-evolution-spectrogram}
\end{figure}

The early-time flow-plate dynamics are found to be largely similar across $m\in[0.0004,33.3162]$ (c.f., figure \ref{fig:mean-overview}(a)), while there is significant variation across $m$ in the long-time dynamics (c.f., figures \ref{fig:mean-overview}(b), (c) and figure \ref{fig:spectral-projection}). To indicate the transitions from the consistent early-time to varied long-time behavior, we show in figure \ref{fig:time-evolution} the full time evolution of the dynamics for emblematic cases across the three regimes, $m= 0.0132,1.44, 5.52$. The time-varying frequency content associated with these signals is shown in figure \ref{fig:time-evolution-spectrogram} as spectrograms.   

For $m=0.0132$ (regime A), the $C_l$ time trace shows small-amplitude oscillations (similar to the rigid-baseline $C_l$ oscillations) until 
$t\approx10$. The $\chi_{mid}$ time trace similarly shows small oscillations until the amplitude grows after $t\approx10$. These dynamics are consistent with the linear behavior described in section \ref{sec:early_time_overview}, which marks the departure from the steady base state as an initial condition. For (roughly) $t\in[10,30]$,
both the $C_l$ and $\chi_{mid}$ time traces show an increase in both the mean as well as amplitude of oscillations. By $t\approx30$ both the $C_l$ and $\chi_{mid}$ curves saturate into limit-cycle oscillations. The time-dependent frequency content can be observed from the spectrogram plots. Figures \ref{fig:time-evolution-spectrogram} (a,d) show that after an initial transient, both the lift and structural displacement evolve at a consistent frequency---very near the reference rigid frequency $f_s$ (indicated by dashed-dotted lines in figure \ref{fig:time-evolution-spectrogram}) (a,d) ---for all time.

For $m=1.44$ (regime B), the small-amplitude linear dynamics persists until $t\approx 10.5$. After $t\approx10.5$, nonlinear effects appear and there is a departure in the mean lift and deflections, until $t\approx36$. The $C_l$ time trace has small-amplitude oscillations with a similar frequency content to the lift variation for the rigid case (see the insets from figure \ref{fig:time-evolution} (b) and the barely visible faint signature in the spectrogram figures \ref{fig:time-evolution-spectrogram} (b) near $f_s$). At $t\approx36$, the $C_l$ time trace abruptly shifts to large-amplitude oscillations that are at a lower frequency than the reference rigid case (evident in figure \ref{fig:time-evolution} (b), with the lower frequency quantified in the spectogram plot from figure \ref{fig:time-evolution-spectrogram}(b)). A similar process of early-time behavior driven near the reference rigid lift frequency, transitioning to larger oscillations at a lower frequency, can also be seen in the $\chi_{mid}$ time-trace. Interestingly, the transition to a lower frequency is found to occur at a different time instance for the lift dynamics. The insets in figure \ref{fig:time-evolution} (b,e) demonstrate how the transition to the lower frequency happens at $t\approx22$ for $\chi_{mid}$ as opposed to $t\approx36$ for the lift signal. By $t=40$, figures \ref{fig:time-evolution-spectrogram} (b,e) show that the $C_l$ and $\chi_{mid}$ time-traces synchronize to a common frequency of approximately $0.38$.

 For $m=5.52$ (regime C), the early-time ($t\lesssim 10$) behavior is driven near the frequency of the reference rigid case (c.f., figure \ref{fig:mean-overview} (a)). After $t\approx10$, 
 both the lift and midpoint displacement signals are seen to evolve onto quasi-periodic oscillations that have a low-frequency modulation. The spectrogram for both the $C_l$ and $\chi_{mid}$ signals suggest dominant frequencies near $0.6$ and 
$1.2$, with the $\chi_{mid}$ time trace showing comparatively stronger signatures near frequency $1.2$ (figure \ref{fig:time-evolution-spectrogram} (c,f)). The frequency values $0.6, 1.2$ fall close to $f_s$ (rigid-baseline frequency) and its harmonic.

These initial observations into the various regimes of behavior motivate a more detailed study, performed in the remainder of this article. We focus on addressing the following key unexplained observations in detail.

For $m\in[0.0004,0.26]$ (regime A), the dynamics exhibit a near-constant frequency independent of the structural mass, and closely aligned with the vortex-shedding frequency in the rigid case, $f_s$. The representative case ($m=0.0132$) presented in figure \ref{fig:time-evolution} shows
that this dominant frequency is associated with the limit-cycle dynamics that appears after an initial transient. Moreover, this frequency in the limit cycle is slightly modulated from that at early time due to the nonlinear saturation process associated with an increase in amplitude. The limit cycle saturates onto a frequency which is near but not identical to the limit-cycle oscillation frequency in the reference rigid case. In section \ref{sec:add_mass_detailed} we explore the mechanisms that set this near-constant frequency, and characterize how the plate oscillations coexist with and modulate the underlying vortex-shedding process.

For intermediate mass values considered in this work, $m\in[0.26,2.76]$ (regime B), the dynamics show a decreasing trend in the spectral content with increasing $m$ becoming increasingly well aligned with the first in-vacuo fundamental frequency of the plate. The peak-to-peak excursions in $\chi_{mid}$, $\overline{\chi_{mid}}$ and $\overline{C_l}$ show 
an increasing trend with $m$, while the peak-to-peak $C_l$ excursions show 
an increasing then decreasing trend with $m$. The discrepancy in where the peaks occur for the different quantities suggest a complex fluid-structure interplay where not all quantities are simultaneously in resonance. The representative case ($m=1.44$) demonstrates a shift in frequency content, from the ubiquitous early-time behavior (driven at the frequency of the rigid reference case) to a lower frequency more nearly aligned with the in-vacuo beam natural frequency. This lower frequency first arises (near $t=25$) in the structural dynamics and then manifests itself in the lift signal (near $t=40$), which raises questions about what triggers the shift in structural frequency and what causes the eventual synchrony in the fluid-structure interaction system. 
In section \ref{sec:mode_1_detailed}, we provide a mechanistic reasoning for the trends seen in both the performance and the spectral content quantities through an analysis of the limit-cycle oscillations  within this regime. We also explore the time-evolution in more detail to probe the routes taken by specific $m$ values towards reaching their respective limit-cycle attractors. 

For higher $m\in[2.76,33.32]$ (regime C), we observe that the beam shows higher structural mode shapes in its oscillations. From the $\overline{C_l}$ and its peak-to-peak excursions, we also observe that these higher mode oscillations have a negligible impact on the performance. In section \ref{sec:mode_2_detailed} we probe the reason behind the circumstances that lead to higher structural mode shapes in the plate oscillations. By analyzing the time-evolution of specific $m$ leading up to this regime, we show how there exists a  ``mode-transition'' across the parametric space $m$. By observing the plate-vortex interplay we also provide a reasoning as to why these higher mode oscillations have little impact on the performance metrics.

\subsection{Fluid-induced mass regime}
\label{sec:add_mass_detailed}

\begin{figure}
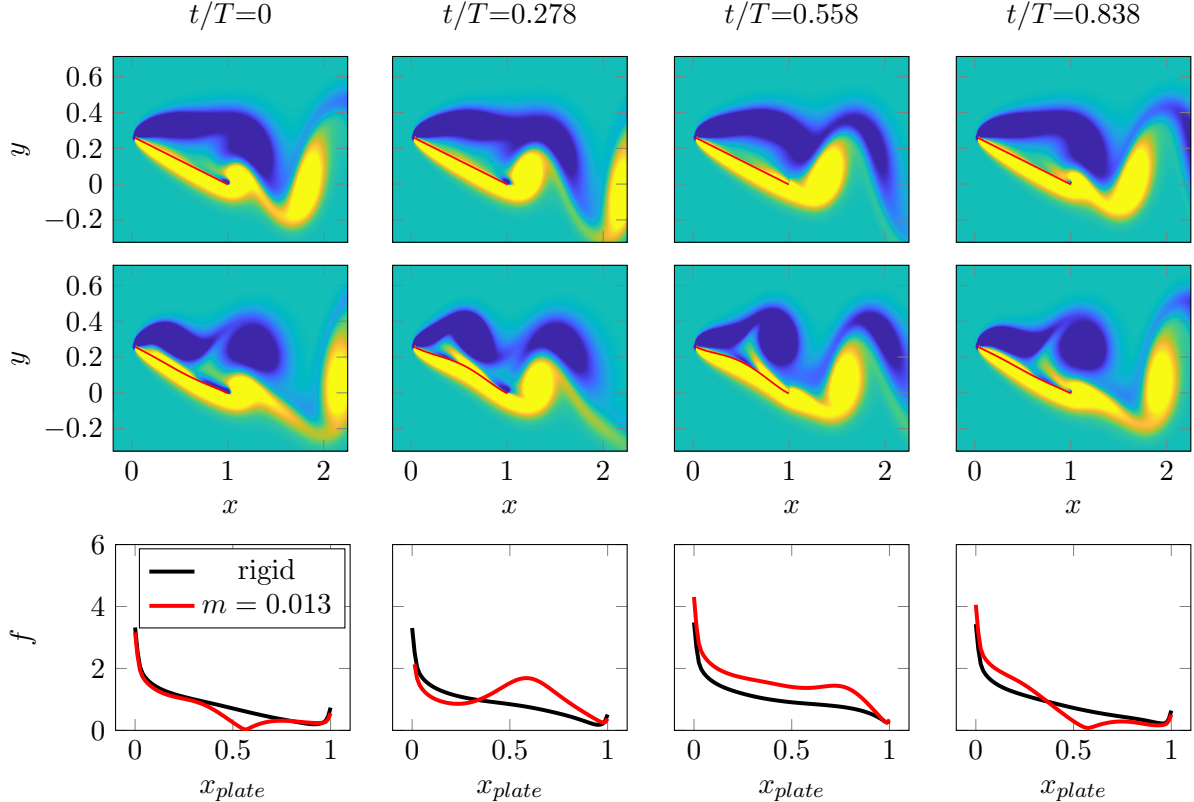


\begin{subfigure}[b]{0.29\textwidth}
\input{figures/overviewfigures_1/rig1label.tex}
\end{subfigure}
\begin{subfigure}[b]{0.22\textwidth}
\input{figures/overviewfigures_1/rig2label.tex}
\end{subfigure}
\begin{subfigure}[b]{0.22\textwidth}
\input{figures/overviewfigures_1/rig3label.tex}
\end{subfigure}
\begin{subfigure}[b]{0.22\textwidth}
\input{figures/overviewfigures_1/rig4label.tex}
\end{subfigure}

\begin{subfigure}[b]{0.29\textwidth}
\input{figures/overviewfigures_1/m0.013_1.tex}
\end{subfigure}
\begin{subfigure}[b]{0.22\textwidth}
\input{figures/overviewfigures_1/m0.013_2.tex}
\end{subfigure}
\begin{subfigure}[b]{0.22\textwidth}
\input{figures/overviewfigures_1/m0.013_3.tex}
\end{subfigure}
\begin{subfigure}[b]{0.22\textwidth}
\input{figures/overviewfigures_1/m0.013_4.tex}
\end{subfigure}

\begin{subfigure}[b]{0.29\textwidth}
\input{figures/overviewfigures_1/m0.013_stress_t0.tex}
\end{subfigure}
\begin{subfigure}[b]{0.22\textwidth}
\input{figures/overviewfigures_1/m0.013_stress_t1.tex}
\end{subfigure}
\begin{subfigure}[b]{0.22\textwidth}
\input{figures/overviewfigures_1/m0.013_stress_t2.tex}
\end{subfigure}
\begin{subfigure}[b]{0.22\textwidth}
\input{figures/overviewfigures_1/m0.013_stress_t3.tex}
\end{subfigure}

\caption{ Vorticity snapshots sampled at specific time instances within a vortex-shedding cycle for the rigid-baseline case (top row) and $m=0.013$ (middle row). Bottom row: Magnitude of the resultant surface stress ($f=\sqrt{(f_x)^2+(f_y)^2}$) acting along the length of the plate, at each instance.}
\label{fig:snapshots}
\end{figure}

\begin{figure}
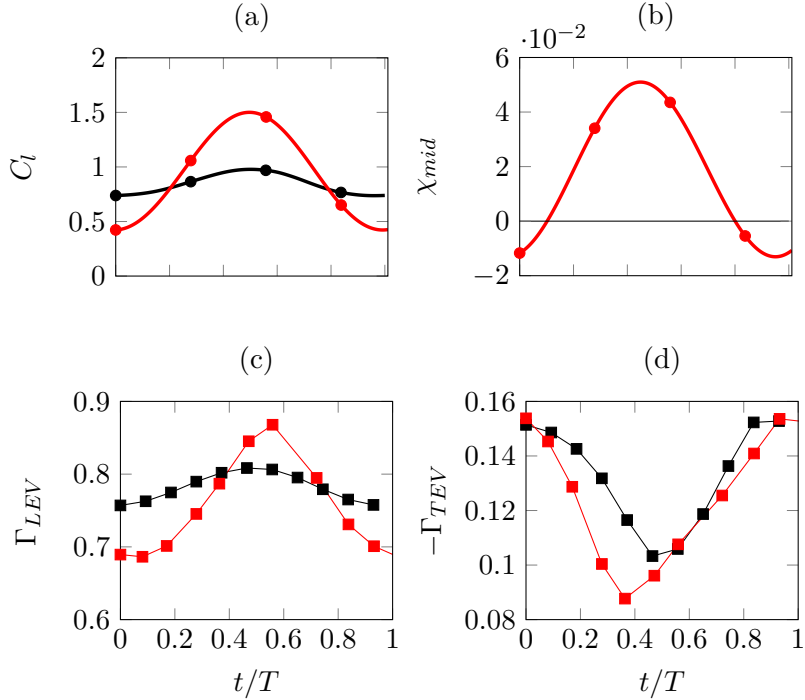

    \centering
    \begin{subfigure}[b]{0.32\textwidth}
\input{figures/addedmassfigures/Fig10_a.tex}
    \end{subfigure}
    \begin{subfigure}[b]{0.32\textwidth}
\input{figures/addedmassfigures/Fig10_b.tex}
    \end{subfigure}
    
    \begin{subfigure}[b]{0.32\textwidth}
\input{figures/addedmassfigures/Fig10_c.tex}
    \end{subfigure}
    \begin{subfigure}[b]{0.31\textwidth}
\input{figures/addedmassfigures/Fig10_d.tex}
\end{subfigure}

    \caption{(a): Time variation of the mean lift-coefficient ($C_l$) over a vortex-shedding cycle, for rigid (black-) and $m=0.013$ (red-solid) cases. The markers indicate the time instances for when the vorticity-snapshots are shown in figure \ref{fig:snapshots}, (b): Time variation of the midpoint displacement of the plate over the considered cycle. The dashed line indicates the zero-displacement shown by the rigid beam, (c): Time variation of the positive circulation strength associated with the LEV over a vortex-shedding cycle. Black markers indicate the corresponding variation seen in the rigid-baseline vortex-shedding cycle. (d):Time variation of the counter-clockwise circulation strength associated with the TEV over a vortex-shedding cycle.}
    \label{fig:markers}
\end{figure}

We compare the limit-cycle behavior of a representative case ($m=0.013$) in regime (A) with the rigid baseline case. To facilitate this discussion, we utilize vorticity snapshots and surface stress information at chosen instances along the limit cycle as shown in figure \ref{fig:snapshots}. We also show in figure \ref{fig:markers} information about the coefficient of lift, midpoint displacement and circulation strength associated with the leading-edge, trailing-edge vortices for both the $m=0.013$ and rigid-baseline case. 

At $t/T=0$, the flexible plate has a slight negative camber compared to the rigid plate. The midpoint of the plate is at $\chi_{mid}\approx-0.01$ (c.f., figure \ref{fig:markers}(b)), and the negative plate camber is associated with an LEV whose spatial structure differs from the rigid case---the LEV associated with the rigid plate spans the entire length of the beam, while for $m=0.013$ the LEV is split into two smaller-sized LEVs (see leftmost figure in middle row of figure \ref{fig:snapshots}). A strong region of negative circulation exists near the fore half of the plate, possibly associated with the negative plate curvature. Figure \ref{fig:snapshots} (bottom row) shows the resultant surface stress acting on the plate at the specified time instances within the cycle. At $t/T=0$, the surface stress magnitude is found to be largely similar in magnitude for both the rigid and flexible plates except around $x_{plate}=0.55$, where the magnitude is significantly smaller for the flexible plate. In figures \ref{fig:markers}(c), (d) we show the time variation of the circulation strength associated with the LEV and the TEV, for both the rigid and $m=0.013$ plates. At $t/T=0$, the TEV circulation strength for $m=0.013$ shows the same magnitude as the rigid-baseline case. However, at the same instance, the $m=0.013$ plate has a lower magnitude in its clock-wise circulation strength associated with the LEV. This lower LEV circulation at $t/T=0$ is associated with a lower $C_l$ as compared to the baseline case (see figure \ref{fig:markers}(a)).

From $t/T=0$ to $t/T=0.278$, both $m=0.013$ and the rigid plates show an increase in $C_l$---with the increase being larger for $m=0.013$. The increase in $C_l$ is associated with an increase in the LEV's circulation strength (see figure \ref{fig:markers}(c)) and a corresponding decrease in the TEV's circulation strength (see figure \ref{fig:markers}(d)). The decrease in the TEV's circulation strength is more prominent for $m=0.013$. During the specified time interval, the $m=0.013$ plate (see figure \ref{fig:markers}(b)) moves upwards from a position of negative camber at $t/T=0$ to a positive camber at $t/T=0.27$. The midpoint of the plate reaches $\chi_{mid}\approx0.03$ (see figure \ref{fig:markers}(b)) above the zero-cambered reference position. 

At $t/T=0.278$, the region of counter-clockwise circulation that was earlier near the fore half of the plate (see the snapshot at $t/T=0$), can now be seen near the plate's position of maximum camber (see second row, second column of figure \ref{fig:snapshots} for visual clarity). Comparing the vorticity contours at $t/T=0.278$, the TEV for $m=0.013$ has advected more downstream in comparison to the baseline case. The slightly faster downstream advection could be associated with the reduced TEV strength seen for $m=0.013$ (see figure \ref{fig:markers}(d)). A comparison between the surface stress distribution between the two cases  
 (see surface stress profile at $t/T=0.278$ in figure \ref{fig:snapshots}) shows that $m=0.013$, on average, has a larger magnitude of surface stress acting across its length. The $m=0.013$ plate shows a peak at $x_{plate}\approx0.6$ as opposed to the rigid plate that shows a flat distribution. The peaks in the surface stress reveal signatures of LEV/TEV and could indicate the proximity between the plate and prominent vortex structures.

From $t/T=0.278$ to $t/T=0.558$, the flexible plate continues its upward trajectory. From figure \ref{fig:markers}, the $C_l$, LEV circulation strength and $\chi_{mid}$ are all shown to increase in magnitude. At $t/T=0.558$, the LEV strength is at its maximum within the cycle. The $\Gamma_{LEV}$ for $m=0.013$ is slightly larger in magnitude compared to the rigid plate. We hypothesize that this could be due to the flexible plate's upward motion leading up to this instant, aiding the LEV roll-up process that is in the same clockwise sense. At $t/T=0.558$, the flexible plate's position is in close proximity to the LEV (c.f., second row, third column in figure \ref{fig:snapshots}) resulting in larger magnitudes of surface stress compared to the rigid plate. 

We compare the variation in the TEV's circulation strength until $t/T=0.558$ (see figure \ref{fig:markers}(d)). The rigid-baseline shows its minimum TEV strength at $t/T\approx0.559$, while $m=0.013$ shows its minimum at $t/T\approx0.35$. At $\frac{t}{T}=0.558$, the rigid plate has not yet begun its TEV roll-up, while the TEV formation is already in process for $m=0.013$. Figure \ref{fig:snapshots} (middle row, $t/T=0.558$) shows that the growing TEV occupies an appreciable region near the aft end of the plate. The presence of a newly forming counter-clockwise circulation region near the plate's aft curvature promotes a region of reverse-flow above the flexible plate.

From $t/T\in[0.558,0.838]$, the flexible plate moves downwards from $\chi_{mid}\approx0.04$ to $\chi_{mid}\approx-0.01$ (c.f., figure \ref{fig:markers}(b)).
The circulation strength associated with the LEV decreases in magnitude while the TEV's circulation strength increases (figure \ref{fig:markers}(c), figure \ref{fig:markers}(d)). The time window is also associated with the advection of the LEV (see snapshots at $t/T\in[ 0.558,0.838]$). The plate's downward motion and the LEV's downstream advection result in a slightly lowered magnitude of surface stress distribution at $t/T=0.838$ when compared to $t/T=0.558$ (see figure \ref{fig:snapshots}). The flexible plate shows a large corresponding drop in $C_l$ over this time window. At $t/T=0.838$, a large region of reverse flow exists above the plate and could be attributed to the presence of a strong TEV $(\Gamma_{TEV}\approx0.15)$ (see figure \ref{fig:markers}(d)).

Building on the analysis of the representative case ($m=0.013$), we further probe the driving FSI mechanisms by comparing results across the different $m$ values. For low $m$ values in regime (A),  $m\in[0.0004, 0.26]$, section \ref{sec:early_time_overview}, \ref{sec:long_time_overview} demonstrated that the frequency, mean lift and displacement, and oscillation amplitude of lift and displacement were nearly constant. The constancy in quantities suggests that for sufficiently low structural mass values, the driver in the mass of the fluid-immersed structure is dominated by the flow rather than the plate's actual mass $m$. 

\begin{figure}
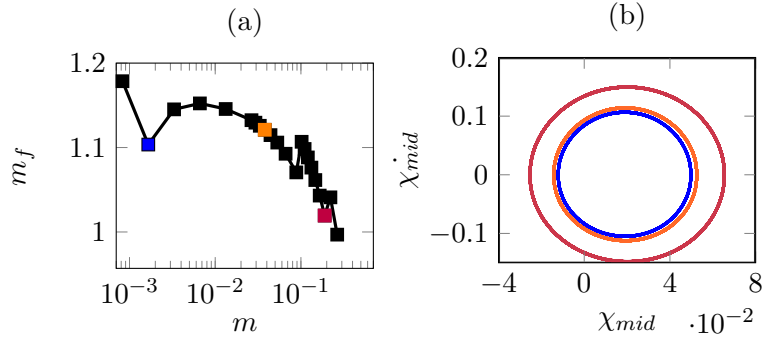

\centering
\begin{subfigure}[b]{0.3\textwidth}
\input{figures/addedmassfigures/Fig11_a.tex}
\end{subfigure}
\begin{subfigure}[b]{0.28\textwidth}
\input{figures/addedmassfigures/Fig11_b.tex}
\end{subfigure}
\caption{(a): Dimensionless added mass coefficient $m_f$ for $m$ in regime A. (b):  Phase portraits associated with the structural midpoint dynamics, for $m=0.0016$ (blue), $m=0.03$ (orange), $m=0.189$ (purple). }
\label{fig:trend}
\end{figure}

In figure \ref{fig:trend}(a), we approximate the dimensionless value of this fluid-induced mass, $m_f$, for each $m$ belonging to regime (A). To approximate this fluid-induced mass, we first compute the dominant frequency associated with the midpoint dynamics of the plate within the FSI system ($\nu^{FSI}$). We then equate this oscillation frequency of the FSI system to a vacuum natural frequency, $\nu^{vacuum}_{1}=(4.73^2/2\pi)\sqrt{(k/(m))}$ \cite{naturalfrequencyeulerbeam} of an equivalent in-vacuo plate (clamped-clamped beam) with a correction that includes the mass of the fluid $m_f$. The associated mass $m+m_f$ (for the fixed stiffness value $k$) that yields this natural frequency is then obtained readily via [$ \nu^{FSI}=(4.73^2/2\pi)\sqrt{(k/(m+m_f))}$]. The fluid-induced mass, $m_f$, may be backed out from the oscillation ``natural'' frequency of the FSI structure and the first natural frequency associated with the in-vacuo plate ($\nu^{vacuum}_1$) as    
\begin{equation}
    m_f=\left(\frac{(\nu^{vacuum}_{1})^2}{(\nu^{FSI})^2}-1\right)m 
\end{equation}

This approach is utilized in place of an analytical form of the fluid-induced mass, which is difficult to obtain given the complex flow dynamics and deforming structural behavior. Note that the computed fluid-loaded mass accounts for added mass, circulatory, and viscous effects. We aim here not to disentangle these various components, but to demonstrate generally that for low values of structural mass $m$ the flow is the driver in the effective mass of the structure immersed in the fluid. 

 In figure \ref{fig:trend}(a), the computed $m_f$ are shown as black markers. For low $m\in[0.0004,0.1]$, $m_f\approx1\gg m$. Thus, across several orders of magnitude of change in $m$, the mass of the structure accounting for fluid effects is $m_f + m \approx m_f$. This near constant mass value (using the fixed stiffness selected for this parameter study) suggests a reason for the similar dynamics across this large range of structure mass values $m$. For $m>0.1$, however, the structural mass $m$ becomes comparable to the flow-induced mass, and the dynamics of the FSI system begin to change as described in sections \ref{sec:long_time_overview}--\ref{sec:overview_summary}. In the other limit of $m$ becoming increasingly large, one expects the structural mass to become an increasingly important driver in the timescales of the structural response (accounting for the periodic loading of vortex shedding, which is always present due to the separated angle of attack).
 
To further probe the differing extents to which the fluid-induced mass drives the coupled dynamics, we focus on three specific cases of $m$ ($m=0.0016, 0.03, 0.189$) in the regime (colored by blue, red and yellow squares in figure \ref{fig:trend}(a)). These three mass values are chosen because they differ by an order of magnitude. Despite this difference in magnitude, the larger value of $m_f$ relative to each $m$ leads us to expect similar detailed dynamics across the three cases (which are also anticipated from their comparable long-time dynamics, c.f., sections \ref{sec:long_time_overview}--\ref{sec:overview_summary}). In what follows, we will highlight the small differences across the cases, indicating the role of flow-induced mass changes where appropriate. It should be kept in mind that these differences are secondary to what is very similar behavior across this wide range of mass values.

For the three considered cases, figure \ref{fig:trend}(b) shows the phase portraits associated with the midpoint displacement and its velocity ($\chi_{mid}$ versus $\dot{\chi}_{mid}$). For the portion of phase space corresponding to  $\dot{\chi}_{mid}>0$, the plate motion is from bottom-to-top with the velocity going to zero at the position of maximum $\chi_{mid}$. The top-to-bottom plate motion corresponds to a negative velocity ($\dot{\chi}_{mid}<0$) as represented by the bottom half of the phase portrait. For all of $m=0.0016, 0.03, 0.189$, the portraits are the same elliptical shape (the curves are scaled versions of one another), with the respective maximum velocity reached at $\chi_ {mid}\approx0.02$. The largest mass value, $m=0.189$, shows the largest trajectory. This result indicates that $m=0.189$ has the largest magnitudes of positive and negative camber at positions of zero velocity. It also demonstrates that $m=0.189$ has the largest velocity when $\chi_{mid}=0$. From $m=0.0016$ to $m=0.189$, there is a monotonic increase in the surface area covered by the $\dot{\chi}_{mid}-\chi_{mid}$ trajectories. This could possibly be due to the slight increase in effect of structural mass relative to the fluid-induced mass, though we emphasize that this effect is of second order: the dynamics are markedly similar across several orders of magnitude of $m$.

\begin{figure}
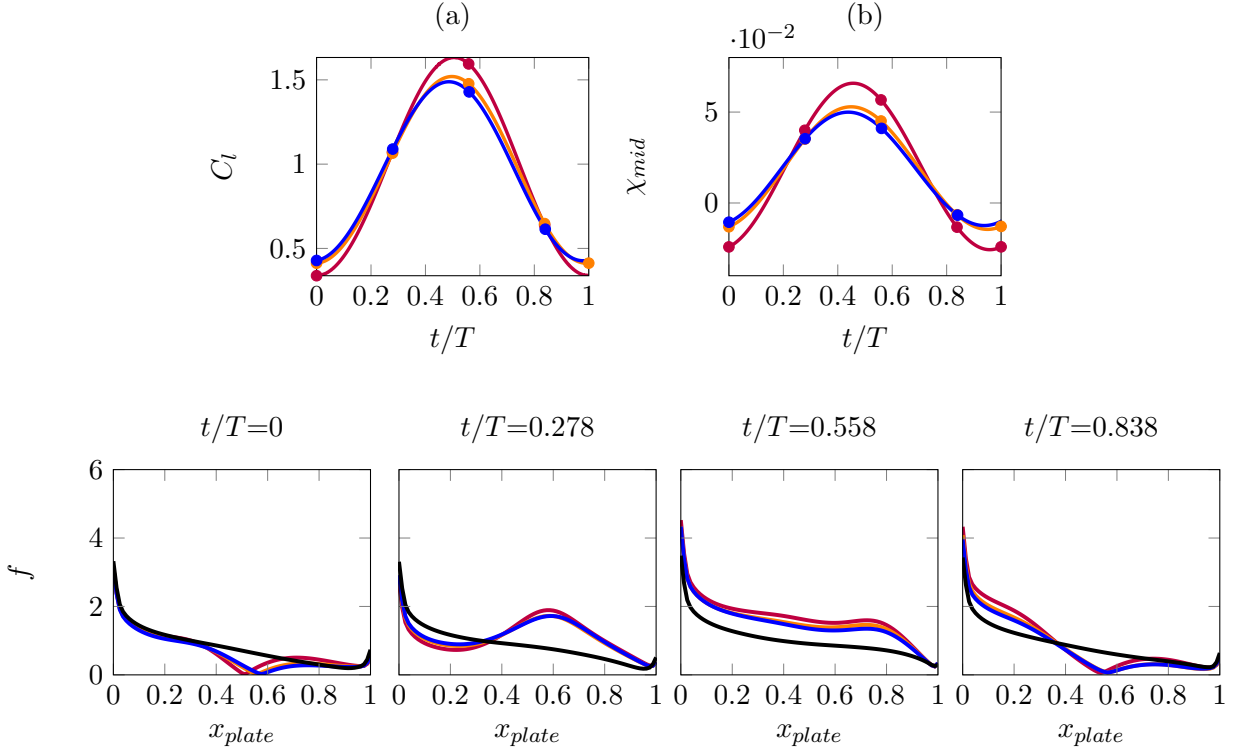

    \centering
\begin{subfigure}[b]{0.33\textwidth}
\input{figures/addedmassfigures/clcyclecomp.tex}
\end{subfigure}
\begin{subfigure}[b]{0.32\textwidth}
\input{figures/addedmassfigures/midptcyclecomp.tex}
\end{subfigure}

\centering
\begin{subfigure}[b]{0.3\textwidth}
\input{figures/addedmassfigures/Fig12_surface_0.tex}
\end{subfigure}
\begin{subfigure}[b]{0.22\textwidth}
\input{figures/addedmassfigures/Fig12_surface_1.tex}
\end{subfigure}
\begin{subfigure}[b]{0.22\textwidth}
\input{figures/addedmassfigures/Fig12_surface_2.tex}
\end{subfigure}
\begin{subfigure}[b]{0.22\textwidth}
\input{figures/addedmassfigures/Fig12_surface_3.tex}
\end{subfigure}

\caption{ (a,b): Time trace plots of $C_l$ and $\chi_{mid}$ over a vortex-shedding cycle, for $m=0.001, 0.03, 0.189$. Bottom row: Resultant surface stress ($f$) distribution along the length of the plate for $m=0.001, 0.037, 0.19$, at instances $t/T = 0, 0.278, 0.556, 0.838$ in a vortex-shedding cycle. For all plots, the colors are the same as described in figure \ref{fig:trend}.}
\label{fig:cyclecomp}
\end{figure}

\begin{figure}
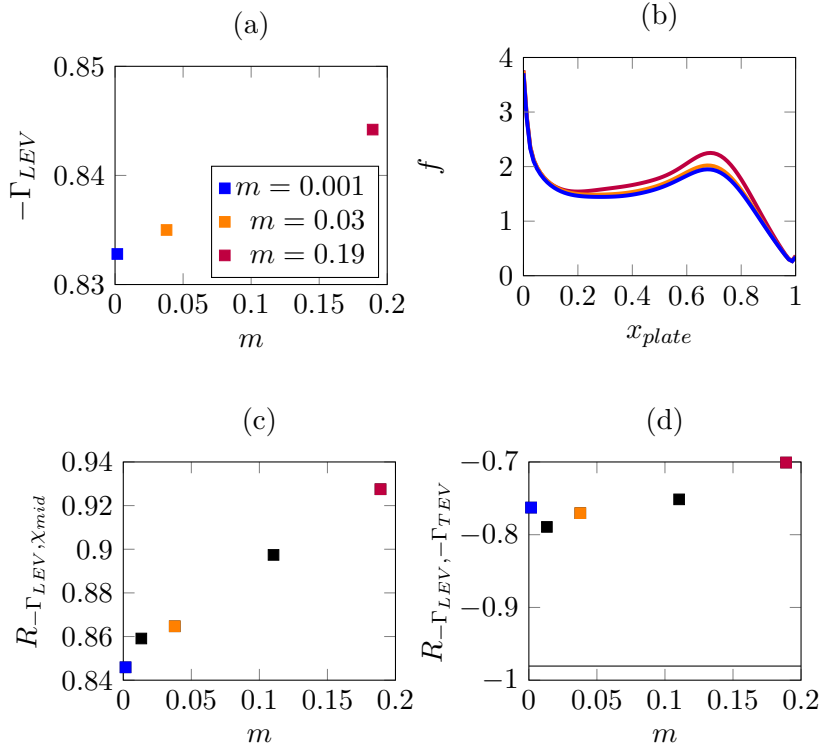

    \centering
    \begin{subfigure}[b]{0.32\textwidth}
\input{figures/addedmassfigures/Fig13_a.tex}
\end{subfigure}
\begin{subfigure}[b]{0.33\textwidth}
\input{figures/addedmassfigures/Fig13_b.tex}
\end{subfigure}

\begin{subfigure}[b]{0.32\textwidth}
\input{figures/addedmassfigures/Fig13_c.tex}
\end{subfigure}
\begin{subfigure}[b]{0.32\textwidth}
\input{figures/addedmassfigures/Fig13_d.tex}
\end{subfigure}

\caption{ (a): Circulation strength associated with the leading-edge vortex at the instance when $\chi_{mid}$ is maximum in the cycle, (b): Resultant surface stress distribution along the length of the plate ($f$) computed at time instance of maximum $\chi_{mid}$, (c): Correlation coefficient between the time varying beam midpoint displacement ($\chi_{mid}$) and the LEV circulation strength ($R_{-\Gamma_{LEV}, \chi_{mid}}$) over a vortex-shedding cycle for $m=0.001,0.03,0.19$, (d): Correlation coefficient between the circulation strength associated with the LEV and the TEV ($R_{-\Gamma_{LEV}, \Gamma_{TEV}}$) over a vortex-shedding cycle for the three considered $m$. Also shown are the values associated with $m=0.01, 0.11$ (black markers) and the rigid-baseline case (solid horizontal line)}
\label{fig:correlation}
\end{figure}

We now assess the impact of these structural dynamics on the flow and aerodynamic performance of the three mass values, $m=0.0016,$ $0.037,$ $0.189$. Figure \ref{fig:cyclecomp} shows the variations associated with $C_l$ (a) and $\chi_{mid}$ (b) over one vortex-shedding cycle. To indicate the evolution of the surface stress throughout the period, the magnitude of the resultant surface stress distribution  at equally spaced time instances within a cycle (indicated by the markers) is shown in the bottom row of figure \ref{fig:cyclecomp}. The surface stress distribution and clockwise circulation strength at the instant of the plate's maximum positive camber are shown in figure \ref{fig:correlation} (a), (b) respectively. Since the flexible plate dynamics affect the timing between the formation and advection processes associated with the LEV and TEV, figures \ref{fig:correlation}(c),(d) show the correlation coefficient between the LEV's circulation strength variation and the midpoint structural displacement, and between the LEV and TEV circulation strengths (respectively).

At $t/T=0$, all the three masses show a dip in the surface stress near $x_{plate}\approx0.5$ relative to the reference rigid case (instance $t/T=0$ in the bottom row of figure \ref{fig:cyclecomp}). The surface stress profile for $m=0.189$ is visibly distinct from the ones seen for $m=0.001, 0.03$. $m=0.189$ shows a slightly more negative camber compared to the other two plates at this instant (see $t/T=0$ in figure \ref{fig:cyclecomp}(b)). From $t/T\in[0,0.278]$, all the three  plates moves upward. At $t/T=0.278$, all the plates show $\chi_{mid}\approx0.04$ (see figure \ref{fig:cyclecomp}(b)), despite $m=0.189$ starting from a slightly higher negative camber at $t/T=0$. This could be associated with the large plate velocities of $m=0.189$ as noted in figure \ref{fig:trend}(b). At $t/T=0.278$, the surface stress profiles in figure \ref{fig:cyclecomp} show that $m=0.189$ has a slightly larger peak near $x_{plate}\approx0.6$ as compared to the other two cases.

For $t/T\in[0.278,0.558]$, the $C_l, \chi_{mid}$ show an increase for all the three cases (figure \ref{fig:cyclecomp}(a), (b)). At $t/T=0.558$, $m=0.189$ shows the largest $C_l, \chi_{mid}$ among the three cases. The corresponding surface stress distribution across the length of the $m=0.189$ plate is also visibly higher (see $t/T=0.558$ in figure \ref{fig:cyclecomp}). 

All the three considered $m$ values reach their respective positions of maximum camber at around $t/T\approx0.5$. At this instance, the circulation strength associated with the LEV is largest for $m=0.189$, while it is the least for $m=0.001$ (see figure \ref{fig:correlation}(a)). At this position, $m=0.189$ also shows a visibly higher magnitude of surface stress across its length (figure \ref{fig:correlation}(b)). A possible reason behind these observations could be that for $m=0.189$, the plate starts its upward motion from a larger negative camber than the other cases, aiding the roll-up process of the LEV in the same clockwise sense.  The larger positive camber of $m=0.189$ also implies that the plate is closer to the low pressure region associated with the LEV, therefore resulting in a higher magnitude of surface stress on the plate.

From $t/T\in[0.558,0.838]$ all the plates move downwards (figure \ref{fig:cyclecomp}(b)) reaching an approximately similar camber at $t/T=0.838$. At $t/T=0.838$, the $C_l$ is also approximately the same for all three cases. The rate of decrease in $C_l$ from $t/T\in[0.559,0.838]$ is largest for $m=0.189$. This is consistent with the $m=0.189$ plate starting its downward motion from a higher positive camber, and with a greater downward velocity. The downward trajectory could benefit the formation of the TEV and promote a region of reverse flow, and thereby larger detriments in lift. 

To indicate the effect of the structural dynamics on the phase relation between the LEV and TEV, the correlation coefficient between $-\Gamma_{LEV}$ and $\Gamma_{TEV}$ is provided in figure \ref{fig:correlation}(d). The value is shown for five cases of increasing $m$, and the horizontal solid line indicates the value for the rigid-baseline (for which $R_{-\Gamma_{LEV},\Gamma_{TEV}}\approx-0.98$). All flexible cases ($m=0.0016,0.0132,0.03,0.189$) show a correlation coefficient that is less negative ($R_{-\Gamma_{LEV},\Gamma_{TEV}}\in[-0.7,-0.8]$). A slight increasing trend is noted from $m=0.0016$ to $m=0.189$. 

We provide reasoning as to why the flexible plate shows a different LEV-TEV interplay compared to the baseline rigid case. The plate oscillations induce TEV formation even before the LEV has completely formed. This can be seen for $m=0.013$ (red curves in figures \ref{fig:markers}(c), \ref{fig:markers}(d)), where there is a considerable duration of overlap between LEV formation and TEV formation (in the baseline case, TEV formation begins only when the LEV begins to advect away from the plate (see black curves in figures \ref{fig:markers}(c), \ref{fig:markers}(d)). The flexible plate has an attached region of counter-clockwise circulation that affects LEV formation during the plate's upward stroke. The upward motion of the plate leads to premature TEV circulation while simultaneously resulting in a prolonged formation phase for the LEV. 

The slight increasing trend in the correlation from $m\in[0.0016,0.189]$ could be associated with the frequency of the plate oscillations. This frequency decreases with increasing mass value $m$, as the plate mass contributes more to the fluid-structure mass and thus resulting in a lower natural frequency of the fluid-embedded plate. Consistent with this lower frequency are larger plate strokes (figure \ref{fig:trend}(b)). The longer time duration possibly leads to an extended formation phase for the LEV, through the same arguments that are made in the preceding paragraph. A greater time duration of overlap between the LEV and TEV formation phases is more likely, possibly resulting in a slightly reduced out-of-phase correlation. It is unsurprising that the difference in LEV-TEV correlation is so small, as for this regime all dynamics are so similar. It is again the second order effects that drive the small differences across this regime.

 Figure \ref{fig:correlation}(c) shows the correlation between the LEV's circulation strength variation and the plate's midpoint displacement,  $\chi_{mid}$. For all considered $m$ values, the plate motion is largely in phase with the LEV, as seen from the positive correlation values. The LEV formation is more synchronized with the plate's upward motion from negative camber to positive camber, and it follows that the LEV's advection/TEV's formation is subsequently more synchronized with the plate's downward motion from positive to negative camber. A slightly increasing trend is seen in the correlation coefficient from $m=0.0016$ ($R_{-\Gamma_{LEV},\chi_{mid}}\approx0.84$) to $m=0.189$ ($R_{-\Gamma_{LEV},\chi_{mid}}\approx0.92$). The mechanistic reasoning behind an increasing $R_{-\Gamma_{LEV},\chi_{mid}}$ could be associated with the increasing magnitude of plate oscillations from $m=0.0016-0.189$. As the plate oscillations become larger with increasing mass across $m\in[0.0016,0.189]$, the frequency associated with the oscillations decreases. This decreased frequency could in turn be associated with an increasing alignment between LEV formation/advection and the plate's oscillations. At $m=0.19$, the phase relation between the vortex processes and the plate motion is largely synchronous, with a high $R_{-\Gamma_{LEV},\chi_{mid}}\approx0.93$.

\subsection{Structure mode 1 regime}
\label{sec:mode_1_detailed}

\begin{figure}
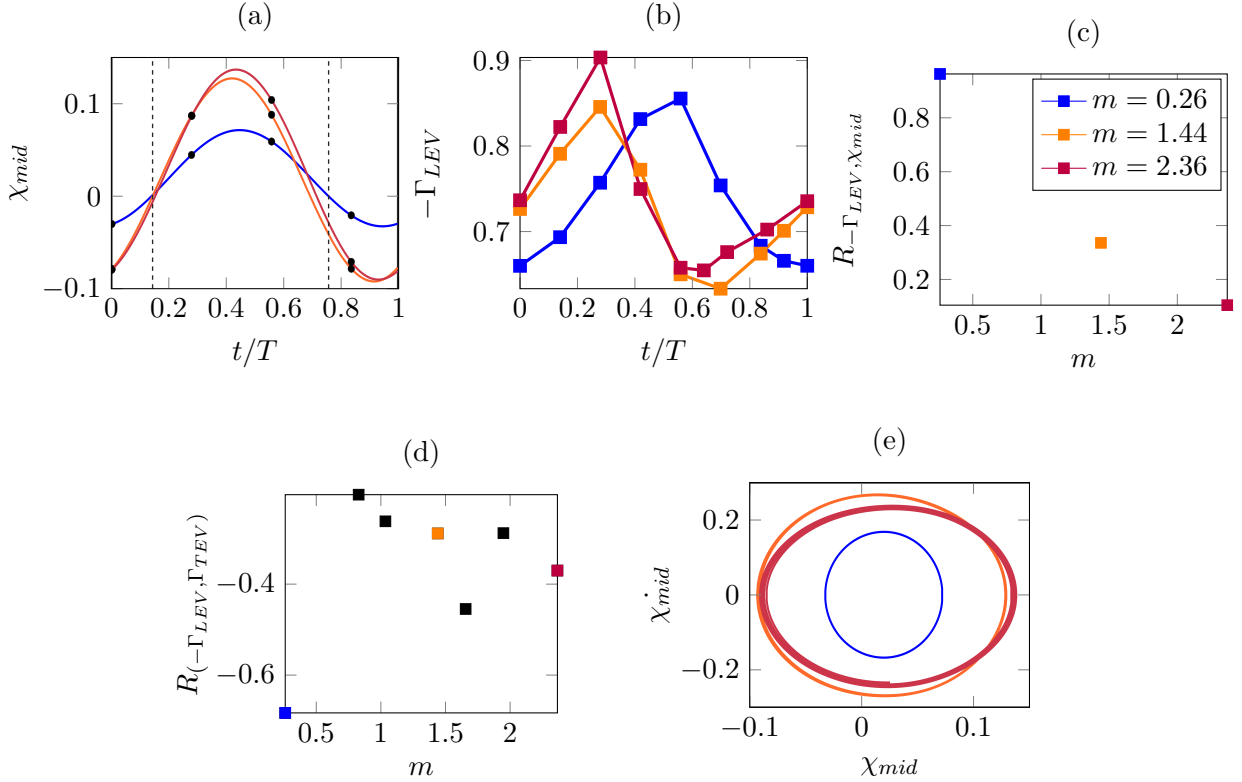

\begin{subfigure}[b]{0.32\textwidth}
\input{figures/mode1figures/Fig14a.tex}
\end{subfigure}
\begin{subfigure}[b]{0.33\textwidth}
\input{figures/mode1figures/Fig14b.tex}
\end{subfigure}
\begin{subfigure}[b]{0.31\textwidth}
\input{figures/mode1figures/Fig14c.tex}
\end{subfigure}

\begin{subfigure}[b]{0.37\textwidth}
\input{figures/mode1figures/Fig14d.tex}
\end{subfigure}
\begin{subfigure}[b]{0.37\textwidth}
\input{figures/mode1figures/Fig14e.tex}
\end{subfigure}

\caption{ (a): Variation of the midpoint displacement of the plate ($\chi_{mid}$) over the cycle, (b): Variation of the circulation strength associated with the leading-edge vortex $\Gamma_{LEV}$ over a lift cycle for $m=0.26, 1.44, 2.36$, (c): Correlation coefficient ($R_{-\Gamma_{LEV}, \chi_{mid}}$) for $m=0.26, 1.44, 2.36$, (d): Correlation coefficient $R_{-\Gamma_{LEV}, \Gamma_{TEV}}$ for $m=0.26, 1.44, 2.36$, (e): Phase portraits associated with the midpoint displacement dynamics, for $m=0.26, 1.44, 2.36$.}
\label{fig:mode1-phase}
\end{figure}

We consider the limit-cycle behavior for  regime B from figure \ref{fig:spectral-projection}. Recall that this regime has structural dynamics dominated by mode one, and is associated with  the structural mass playing a significant role (relative to the fluid-induced mass) in setting the dynamics. To facilitate the discussion, we study in detail three representative cases across the parametric space $m$ in regime B:  $m=0.26, 1.44, 2.36$. 

\begin{figure}
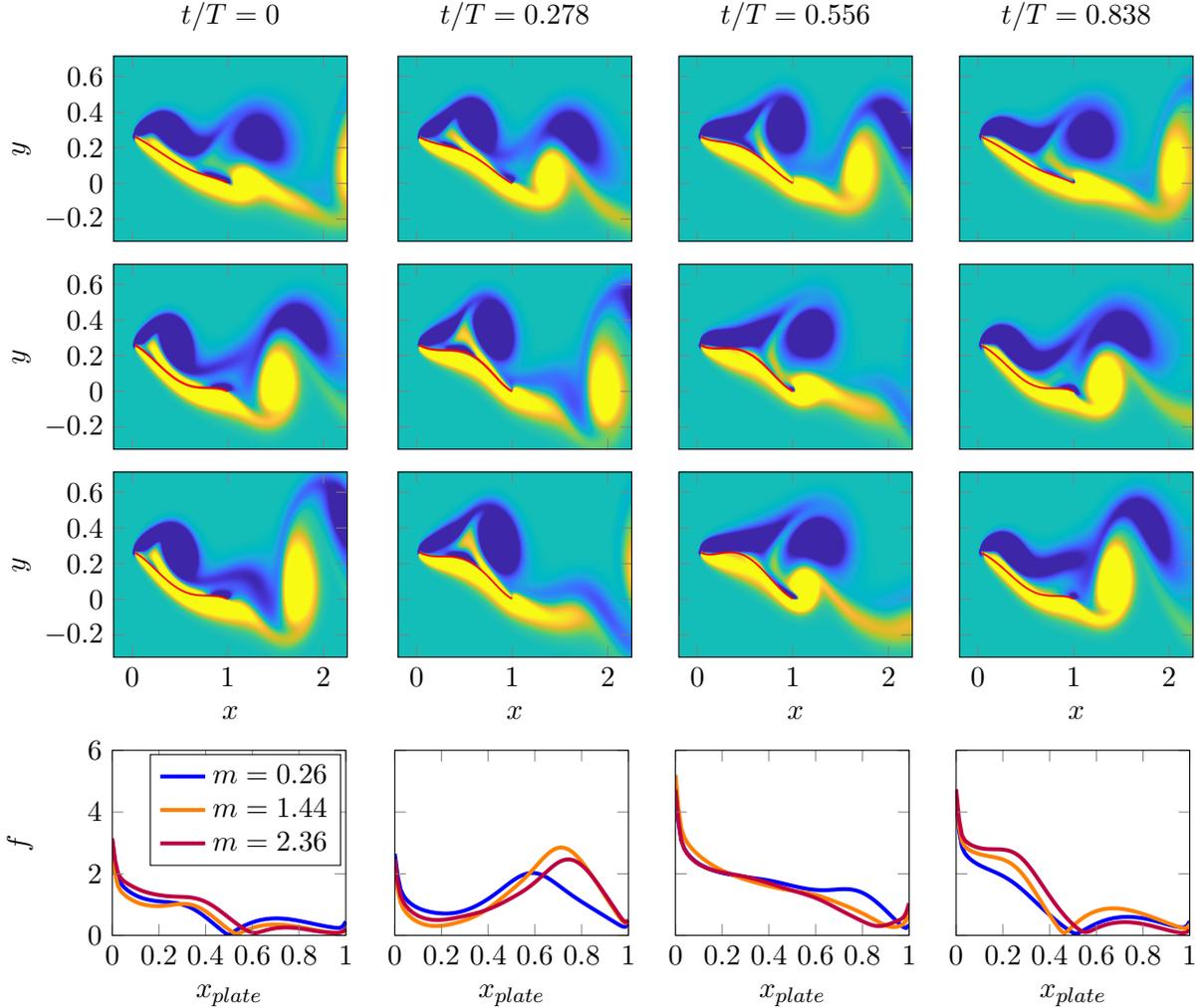

\begin{subfigure}[b]{0.3\textwidth}
\input{figures/mode1figures/Fig15_11.tex}
\end{subfigure}
\begin{subfigure}[b]{0.22\textwidth}
\input{figures/mode1figures/Fig15_12.tex}
\end{subfigure}
\begin{subfigure}[b]{0.22\textwidth}
\input{figures/mode1figures/Fig15_13.tex}
\end{subfigure}
\begin{subfigure}[b]{0.22\textwidth}
\input{figures/mode1figures/Fig15_14.tex}
\end{subfigure}

\begin{subfigure}[b]{0.3\textwidth}
\input{figures/mode1figures/Fig15_21.tex}
\end{subfigure}
\begin{subfigure}[b]{0.22\textwidth}
\input{figures/mode1figures/Fig15_22.tex}
\end{subfigure}
\begin{subfigure}[b]{0.22\textwidth}
\input{figures/mode1figures/Fig15_23.tex}
\end{subfigure}
\begin{subfigure}[b]{0.22\textwidth}
\input{figures/mode1figures/Fig15_24.tex}
\end{subfigure}

\begin{subfigure}[b]{0.3\textwidth}
\input{figures/mode1figures/Fig15_31.tex}
\end{subfigure}
\begin{subfigure}[b]{0.22\textwidth}
\input{figures/mode1figures/Fig15_32.tex}
\end{subfigure}
\begin{subfigure}[b]{0.22\textwidth}
\input{figures/mode1figures/Fig15_33.tex}
\end{subfigure}
\begin{subfigure}[b]{0.22\textwidth}
\input{figures/mode1figures/Fig15_34.tex}
\end{subfigure}

\begin{subfigure}[b]{0.3\textwidth}
\input{figures/mode1figures/Fig15_41.tex}
\end{subfigure}
\begin{subfigure}[b]{0.22\textwidth}
\input{figures/mode1figures/Fig15_42.tex}
\end{subfigure}
\begin{subfigure}[b]{0.22\textwidth}
\input{figures/mode1figures/Fig15_43.tex}
\end{subfigure}
\begin{subfigure}[b]{0.22\textwidth}
\input{figures/mode1figures/Fig15_44.tex}
\end{subfigure}

\caption{ Vorticity snapshots sampled at specified time instances (as indicated by the black markers in figure \ref{fig:mode1-phase}(b)) over a $\chi_{mid}$ cycle. Snapshots for $m=0.26$ (first row), $m=1.44$ (second row) and $m=2.36$ (third row) are shown. For the three considered $m$, the bottommost row shows the magnitude of the resultant
surface stress ($f$) acting along the length of the plate, at each instance. }
\label{fig:mode1longtimesnapshots}
\end{figure}

For these representative cases, figures \ref{fig:mode1-phase}-- \ref{fig:mode1longtimesnapshots} show information about the vortex-plate interplay over one cycle. For each considered case, the plate originally starts with a slight negative camber at $t/T=0$, reaches its maximum camber at around $t/T\approx0.4$ and completes its full stroke by $t/T\approx0.9$ (see $\chi_{mid}$ plot in figure \ref{fig:mode1-phase}(a)).

Figure \ref{fig:mode1-phase}(b) shows the circulation strength variation associated with the leading-edge vortex. For $m=0.26$, the circulation strength associated with the LEV increases in magnitude from $t/T\in[0,0.5]$ (i.e., this is the LEV formation phase) and shows a subsequent decrease from $t/T\in[0.5,1]$ (i.e., LEV advection phase). For $m=0.26$, visually, figures \ref{fig:mode1-phase}(a) and (b) indicate that $\Gamma_{LEV}$ and $\chi_{mid}$ evolve nearly in phase throughout a cycle. The variation in the LEV's circulation strength happens differently for $m=1.44, 2.36$ in comparison to $m=0.26$. The formation phase of the LEV (for both $m=1.44,2.36$) takes place until $t/T\approx0.23$, and the advection phase takes place from $t/T\approx\in [0.23-0.6]$. From $t/T\in[0.6,1]$, there is another formation phase for the LEV. When comparing these variations with the corresponding $\chi_{mid}$ oscillations, the plate's upward motion is partially associated with both LEV formation and advection.

The phase relation between $\chi_{mid}$ and $\Gamma_{LEV}$ is explicitly computed as a correlation coefficient $R_{-\Gamma_{LEV},\chi_{mid}}$ in figure \ref{fig:mode1-phase}(c). Figure \ref{fig:mode1-phase}(d) compares the correlation between $\Gamma_{LEV}$ and $\Gamma_{TEV}$ ($R_{-\Gamma_{LEV},\Gamma_{TEV}}$) for the three cases. A high value of $R_{-\Gamma_{LEV},\chi_{mid}}\approx0.9$ is seen for $m=0.26$, confirming an in-phase relationship between the signals. $R_{-\Gamma_{LEV},\chi_{mid}}$ is found to be comparatively lower for $m=1.44$ ($\approx0.3$) and $m=2.36$ ($\approx0.1$) than for $m=0.26$ ($\approx0.9$).  While both $m=1.44, 2.36$ show $R_{-\Gamma_{LEV},\Gamma_{TEV}}>-0.4$,  $m=0.26$ shows a high out-of-phase value of $R_{-\Gamma_{LEV},\Gamma_{TEV}}\approx-0.65$. 

To contrast the plate dynamics for $m=0.26, 1.44, 2.36$ figure \ref{fig:mode1-phase}(e) shows the phase portraits associated with the oscillations. $m=0.26$ shows the smallest curve, indicating that it has the smallest positive/negative camber and velocity. The higher masses ($m=1.44,2.36$) have larger displacements and velocities. Moreover, the phase portrait for each mass is found to slightly differ in shape/trajectory. This observation is in contrast to the fluid-induced mass regime where the structural dynamics where scaled variants of one another. This change in shape again speaks to the driving effect of the structural properties in the structure mode one regime compared with the fluid-induced mass regime.

For the three cases ($m=0.26, 1.44, 2.36$), figure \ref{fig:mode1longtimesnapshots} shows vorticity snapshots and the surface stress distribution along the plate, sampled at equally spaced time instances within the cycle. At $t/T=0$, the plates with $m=2.366$ and $m=1.44$ (second and third row) have a higher negative camber when compared to $m=0.26$. At this instance, both $m=1.44$ and $m=2.36$ have a larger magnitude of clockwise circulation strength above the plate (c.f., figure \ref{fig:mode1-phase} (b). The signatures of the strong clockwise circulation strength is reflected in the surface stress profiles: $m=1.44, 2.36$ show a comparatively higher magnitude in the surface stress across the fore-half of the plate ($x_{plate}\in[0,0.5]$). From $t/T\in[0,0.278]$, the plates move upwards. The upward motion is associated with formation (roll up) of the LEV. At $t/T=0.27$, the LEV's circulation strength is highest for $m=2.36$ (c.f., figure \ref{fig:mode1-phase}(b)). The increased LEV strength could be connected to the plate having a larger magnitude of positive camber for this mass value. During the plate's upward motion, an attached counter-clockwise circulation is visually seen to exist on the plate. The presence of the counter-clockwise circulation could possibly be aiding the roll-up of the leading-edge vortex via the no-slip constraint. Since $m=2.36$ shows a higher upward velocity (c.f., the phase portrait in figure \ref{fig:mode1-phase}(e)), the roll-up process for the LEV is possibly more enhanced for $m=2.366$. 

 At $t/T=0.278$, the surface stress profile contains a peak on the latter half of the plate (c.f., bottom row of figure \ref{fig:mode1longtimesnapshots}). For each $m$ value, this peak appears at different spatial positions along the length of the plate. While for $m=0.26$ this peak is near $x_{plate}\approx0.6$, for $m=1.44, 2.36$ it is further downstream at $x_{plate}\approx0.7, 0.8$ respectively. The streamwise location of this peak in the surface stress corresponds strongly with the streamwise location of the LEV. For example, the LEV is almost near the trailing edge of the plate for $m=2.366$ and $m=1.44$, whereas its position is only near the plate's midpoint in case of $m=0.26$. 

From $t/T\in[0.278,0.556]$, the plate continues its upward motion for all mass values. It then reaches its maximum camber and begins its downward motion. By $t/T=0.556$, the plate is on its  downward trajectory for all cases. For $m=0.26$, LEV formation takes place until $t/T\approx0.5$, and the LEV has only just begun its advection downstream (see figure \ref{fig:mode1-phase}(b)). In contrast, the snapshots for both $m=1.44$ and $m=2.366$ show that their LEV has already advected downstream significantly by $t/T=0.556$. The difference in the LEV's position is also reflected in the surface stress distribution. While $m=0.26$ shows a small local peak near $x_{plate}\approx0.8$, no local peaks are seen for $m=1.44, 2.36$. Furthermore, both $m=1.44$ and $m=2.36$ show a prominent TEV (visually evident from snapshot) that has already formed to a significant extent by $t/T=0.556$. The counter-clockwise circulation associated with the TEV promotes a strong region of reverse flow in the vicinity of the plate.

From $t/T\in[0.556,0.838]$ the plate continues its downward motion. By $t/T=0.838$, both $m=1.44$ and $m=2.366$ reach positions of appreciable negative camber while $m=0.26$ only reaches a slight negative camber. The same interplay cycle is repeated again from $t/T=0$.

In summary, the representative flexible plates ($m=0.26, 1.44, 2.36$) show subtle differences in their interplay with the vortex-shedding process. For $m=0.26$, plate oscillations are largely in phase with the LEV's formation and advection processes. This synchrony is less prominent for $m=1.44, 2.36$. For $m=2.36$, the correlation between the LEV and plate motion is almost $90^{\circ}$ ($R_{-\Gamma_{LEV}, \chi_{mid}}\approx 0.1$). Also, the frequency of the plate's oscillations is almost the same as $\nu^{vacuum}_1$, the first fundamental in-vacuo frequency associated with the plate. The fact that for $m=2.36$ the FSI frequency matches in-vacuo fundamental frequency, and that the oscillation magnitudes in plate motion and $\overline{C_l}$ are maximal for this case (c.f., section \ref{sec:long_time_overview}), indicate a resonant interplay. Indeed, we argue that the general trends in coefficient of lift from section \ref{sec:long_time_overview} are reflective of a progressive synchronization between flow and structure timescales as mass increases, leading to larger amplitude structural oscillations. Eventually, for increasing structural mass the first structural natural frequency becomes too low for the flow to ``lock on'', and the flow instead interacts with higher structural modes, discussed in the next section.

\begin{figure}
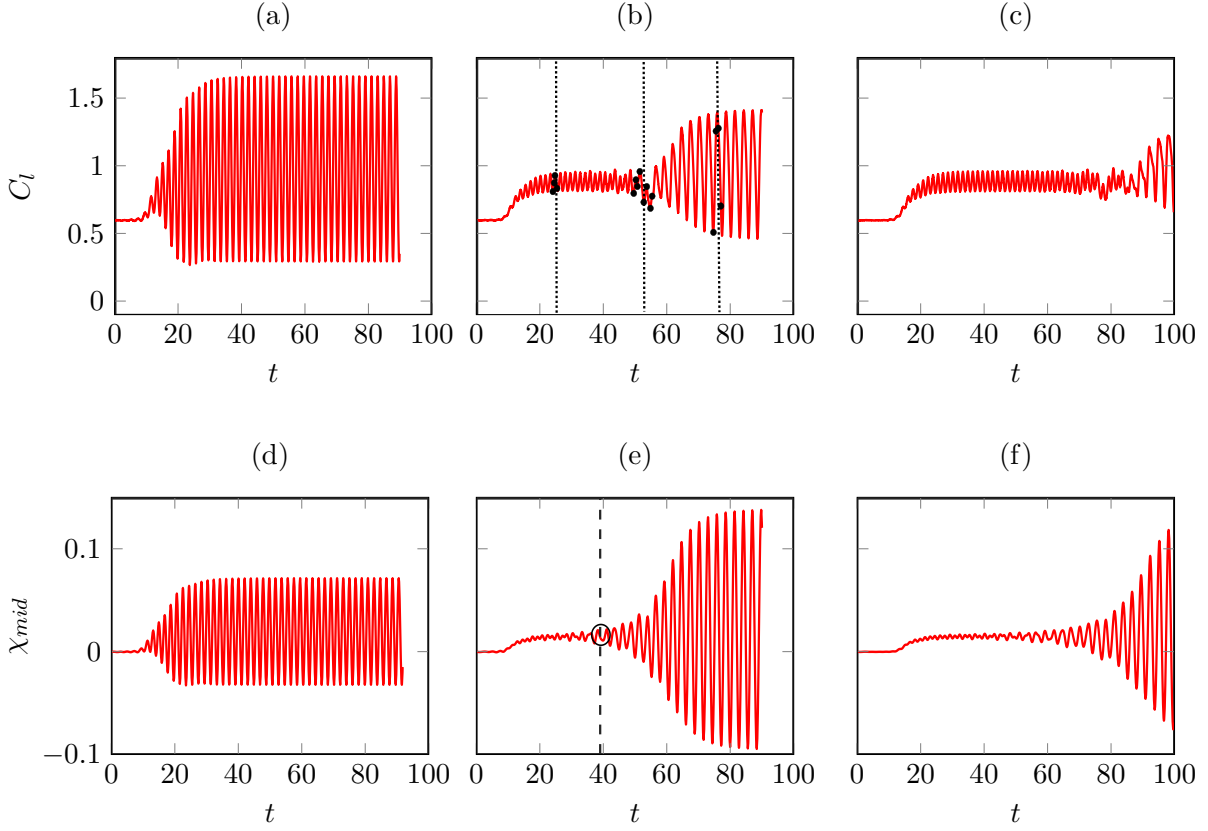

\begin{subfigure}[b]{0.36\textwidth}
\input{figures/mode1figures/Fig16_m0.26.tex}   
\end{subfigure}
\begin{subfigure}[b]{0.3\textwidth}
\input{figures/mode1figures/Fig16_m1.948.tex}  
\end{subfigure}
\begin{subfigure}[b]{0.3\textwidth}
\input{figures/mode1figures/Fig16_m2.36.tex}  
\end{subfigure}

\begin{subfigure}[b]{0.36\textwidth}
\input{figures/mode1figures/Fig16_midpt_m0.26.tex}   
\end{subfigure}
\begin{subfigure}[b]{0.3\textwidth}
\input{figures/mode1figures/Fig16_midpt_m1.948.tex}  
\end{subfigure}
\begin{subfigure}[b]{0.3\textwidth}
\input{figures/mode1figures/Fig16_midpt_m2.36.tex}  
\end{subfigure}

\caption{Time trace plots of the $C_l$ (a-c) and $\chi_{mid}$ (d-f) signals. Each column corresponds
to a different mass value $m$, as: $m = 0.26$ (a,d), $m = 1.948$ (b,e), and $m = 2.36$ (c,f). Dashed lines and markers (b) at $t\approx25$ and $t\approx75$ indicate time instances used to obtain correlation coefficients for figure \ref{fig:mode1freqshift}(c). Dashed lines at $t\approx50$ in (b) and at $t\approx40$ in (e) represent the transition to low-frequency oscillations  }
\label{fig:mode1time}
\end{figure}

To probe the mechanism behind how this ``lock on'' process becomes prominent within the regime, we show in figure \ref{fig:mode1time}  the $C_l$ and $\chi_{mid}$ time trace for $m=0.26, 1.948, 2.366$. For $m=0.26$ (first column), both $C_l$ and $\chi_{mid}$ depart the linear regime at around $t\approx10$ and saturate to limit-cycle oscillations by $t\approx25$. As noted previously in section \ref{sec:long_time_overview}, the limit-cycle oscillations show the same frequency in both the $C_l$ and the $\chi_{mid}$ time traces. The transition to limit-cycle oscillations occurs very quickly after the departure from the linear regime, unlike what is observed for the larger two masses.

The time traces for $m=1.948, 2.36$ show a significant intermediate time window, after the departure from linear behavior from the base state but before the eventual large-amplitude oscillations. Within this intermediate time window, there is a transition in dominant frequency that marks the eventual onset of large-amplitude oscillations. For $m=1.95$ this transition occurs near $t\approx40$ for $\chi_{mid}$ (indicated by the vertical line in figure \ref{fig:mode1time}(e)) and near $t\approx55$ for $C_l$ (indicated by the second vertical line in  figure \ref{fig:mode1time}(b)). For $m=2.36$, the transition occurs near $t\approx60$ for $\chi_{mid}$ and $t\approx80$ for $C_l$. The existence of a switch in frequency for $\chi_{mid}$ before $C_l$ indicates a physical process by which lock on occurs, which we explore in further detail next.

 To probe the nature of the initial transient and how it eventually leads to long-time limit-cycle oscillations, we exclusively focus on the time-trace associated with $m=1.948$. (Conclusions are qualitatively similar for the other mass values within this regime). 
 
\begin{figure}
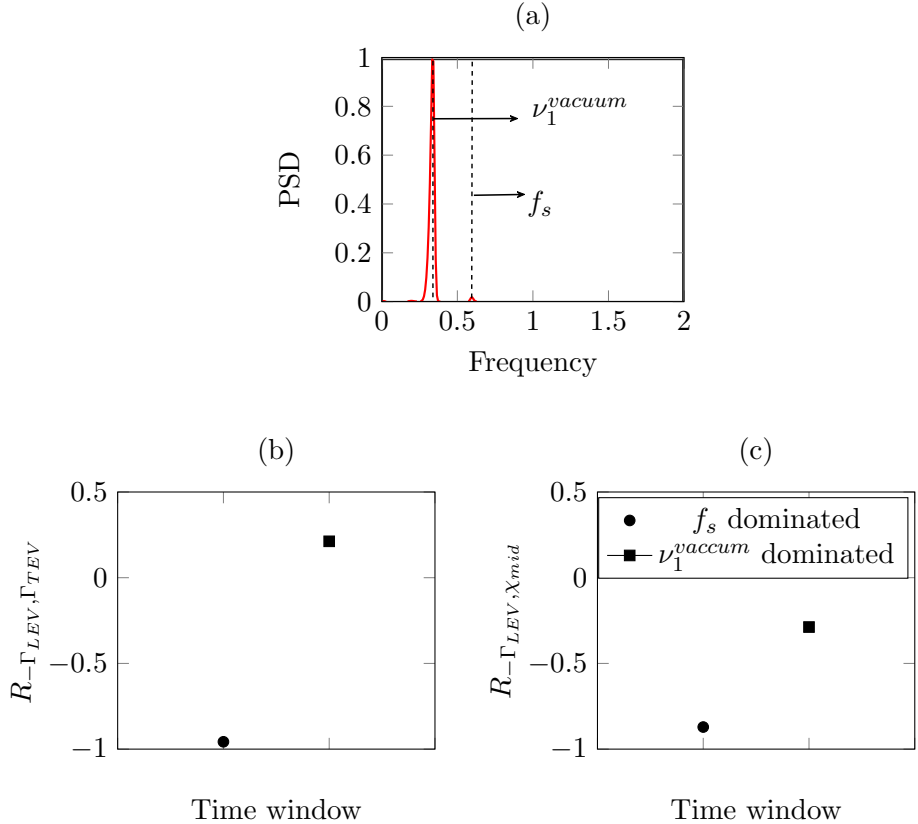


\begin{subfigure}[b]{0.32\textwidth}
\input{figures/mode1figures/Fig17a.tex}  
\end{subfigure}

\begin{subfigure}[b]{0.38\textwidth}
\input{figures/mode1figures/Fig17b.tex}  
\end{subfigure}
\begin{subfigure}[b]{0.38\textwidth}
\input{figures/mode1figures/Fig17c.tex}  
\end{subfigure}

    \caption{ 
    (a): Information about the power spectral density (PSD) of the $\chi_{mid}$ oscillations for $m=1.948$ (also shown are the baseline vortex-shedding frequency $f_s$ and first fundamental frequency $\nu^{vacuum}_1$), (b): Correlation coefficient $R_{-\Gamma_{LEV} \Gamma_{TEV}}$ computed at two different time windows (indicated by two indices) over the time trace. (c): Correlation coefficient $R_{-\Gamma_{LEV}, \chi_{mid}}$ computed at two different time windows (indicated by two indices) over the time trace.  }
    \label{fig:mode1freqshift}
\end{figure}

 Figure \ref{fig:mode1freqshift}(a) shows the power spectral density associated with the $\chi_{mid}$ time trace for $m=1.948$, over the entire time window after the linear dynamics were departed. The dominant frequency seen in the signal is indicated as being equivalent to $\nu^{vacuum}_1$, the first fundamental in-vacuo frequency of the plate. A small peak is also seen near $f_s$, indicating that before the eventual transition to the first (in-vacuo) structural mode, the structure oscillates at the vortex-shedding frequency associated with the rigid plate. The smallness in the peak's magnitude is reflective of the low-amplitude seen in structural oscillations during the transient period. 

We inspect these three distinct behavioral time windows further: small-amplitude oscillations with prominent $f_s$ content, variable-amplitude oscillations transitioning between $f_s$ to $\nu^{vacuum}_1$, and large-amplitude oscillations with prominent $\nu^{vacuum}_1$ content. Note that these distinct time windows are indicated by the different vertical lines in figure \ref{fig:mode1time}(b). The markers surrounding each set of vertical lines indicate the time instances that will be probed in further detail to characterize the dynamics surrounding each regime.

We begin our analysis with the first ($f_s$-dominated) and third ($\nu^{vacuum}_1$-dominated) time windows, as these will form the backdrop for the intermediate second time window. For a vortex-shedding cycle associated with the first and third behavioral time window, figure \ref{fig:mode1freqshift}(b) shows the correlation coefficient between the LEV and TEV circulation strength. The correlation between the plate oscillations and the LEV circulation strength is shown in figure \ref{fig:mode1freqshift}(c). For the first behavioral time window dominated by $f_s$, $R_{-\Gamma_{LEV},\Gamma_{TEV}}\approx-1$ and $R_{\chi_{mid},\Gamma_{LEV}}\approx-1 $. This indicates a near $180^{\circ}$ out-of-phase relation between $\Gamma_{LEV}$ and $\chi_{mid}$. We recall that $R_{-\Gamma_{LEV},\Gamma_{TEV}}=-1$ is a value that is observed in the rigid-baseline case. This phase relation implies that the plate possibly acts a `forced oscillator' that responds to forcing from the baseline vortex-shedding.  At the third time window dominated by $\nu^{vacuum}_1$, $R_{-\Gamma_{LEV},\Gamma_{TEV}}\approx0.2$ and $R_{\chi_{mid},-\Gamma_{LEV}}\approx-0.2$. This phase relation, distinct from that of the $f_s$-dominated time window, indicates the plate is no longer `forced' by the underlying vortex-shedding frequency inherent to the bluff-body flat-plate system. 

These results indicate that for a brief period directly after the dynamically linear regime $(20\lesssim t \lesssim 40)$, for $m=1.948$, the plate's response is primarily governed by forcing from the fluctuating vortex-shedding at frequency $f_s$. The inertia ($m$) of the plate delays the onset of mode one oscillations, and in doing so gives rise to transitory oscillations driven by the baseline vortex-shedding frequency $f_s$. Eventually, the plate is observed to alter its phasing with the vortex-shedding at $t\approx40$ resulting in resonant mode one oscillations.

\begin{figure}
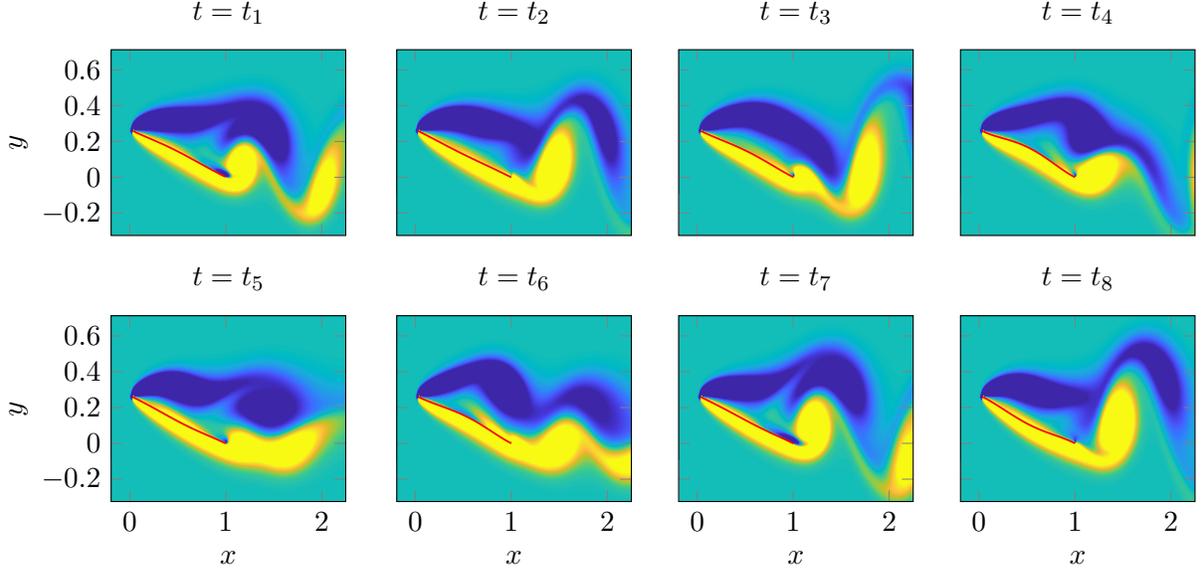

\begin{subfigure}[b]{0.3\textwidth}
\input{figures/mode1figures/Fig18a.tex}
\end{subfigure}
\begin{subfigure}[b]{0.22\textwidth}
 \input{figures/mode1figures/Fig18b.tex} 
\end{subfigure}
\begin{subfigure}[b]{0.22\textwidth}
\input{figures/mode1figures/Fig18c.tex} 
\end{subfigure}
\begin{subfigure}[b]{0.22\textwidth}
\input{figures/mode1figures/Fig18d.tex}  
\end{subfigure}

\begin{subfigure}[b]{0.3\textwidth}
\input{figures/mode1figures/Fig18e.tex}
\end{subfigure}
\begin{subfigure}[b]{0.22\textwidth}
\input{figures/mode1figures/Fig18f.tex}
\end{subfigure}
\begin{subfigure}[b]{0.22\textwidth}
\input{figures/mode1figures/Fig18g.tex}
\end{subfigure}
\begin{subfigure}[b]{0.22\textwidth}
\input{figures/mode1figures/Fig18h.tex}
\end{subfigure}

\caption{Vorticity snapshots sampled at sequential time instances ($t\in[t_1,t_8]$) indicated by black markers in the $C_l$ time trace for $m=1.948$ (see figure \ref{fig:mode1time}(b)).}
\label{fig:mode1freqshiftsnap}
\end{figure}

We now assess the second time window in which there is an onset of switching from $f_s$-dominated to $\nu_1^{vacuum}$-dominated behavior, $t=45-60$.  The marked instances over this window will be labeled $[t_1, t_2, \dots, t_8]$, and are chosen such that they correspond to a local extremum in $C_l$ (peak/trough) in the shedding cycle (see figure \ref{fig:mode1time}(b)). At $t=t_1$, the $C_l$ has attained its minimum within the cycle and the TEV is at its maximum strength (see large region of negative circulation in figure \ref{fig:mode1freqshiftsnap}). From $t=t_1$ to $t=t_2$, the LEV undergoes formation (see roll up of LEV on the top of the plate at $t_2$) and the plate moves slightly downwards. At $t=t_2$ the plate shows a slight negative camber. From instances $t\in[t_2,t_4]$, the plate moves upwards, resulting in a  large positive camber at $t_4$. The plate's upward motion during these instances interrupts the usual formation phase associated with the TEV, via the no-slip constraint. The instances $t_3$ to $t_4$ is associated with an increase in $C_l$. From $t\in[t_4,t_5]$, the plate moves downward. At $t=t_5$, the negative camber associated with the plate results in a larger drop in lift. From $t_5$ to $t_6$, the plate motion is upwards and is in the same sense as the LEV roll-up process. The upward plate motion again intercedes a formation phase associated with the TEV, associated with an increase in $C_l$. From $t_6-t_7$, the plate moves downwards and is again associated with a drop in lift. 

The instances $t\in[t_1,t_7]$ indicate a process whereby the structural motion changes the underlying vortex formation and shedding process. First, one may expect a lowering in the overall structural frequency due to the near-constant lift value associated with the mean camber, induced by the departure from the base state. This near-constant lift offset can be expected to drive structural mode 1 dynamics at their natural frequency, and at these larger structural mass values one might expect this natural frequency to be dominated by the in-vacuo value. The onset of this lower structural frequency is delayed as mentioned above, absent at early times $(20\lesssim t \lesssim 40)$ because the significant structural inertia drives a response to the underlying vortex-shedding content present even before the mean camber arises. After a relatively long time, $t\approx40$, the mode 1 oscillations at their natural frequency begin to manifest and alter the vortex-shedding process as described above, interrupting the formation procedure observed for rigid cases and inducing a new larger-amplitude process where vortex shedding is locked on to the plate motion, at $\nu^{vacuum}_1< f_s$. The lock-on process is complete at $t_8$ when the baseline-like formation phase of the TEV is fully mitigated by plate’s upward motion. The dynamics after $t_8$ have a common frequency in both $\chi_{mid}$ and $C_l$.

\subsection{Higher structural mode regime}
\label{sec:mode_2_detailed}

\begin{figure}
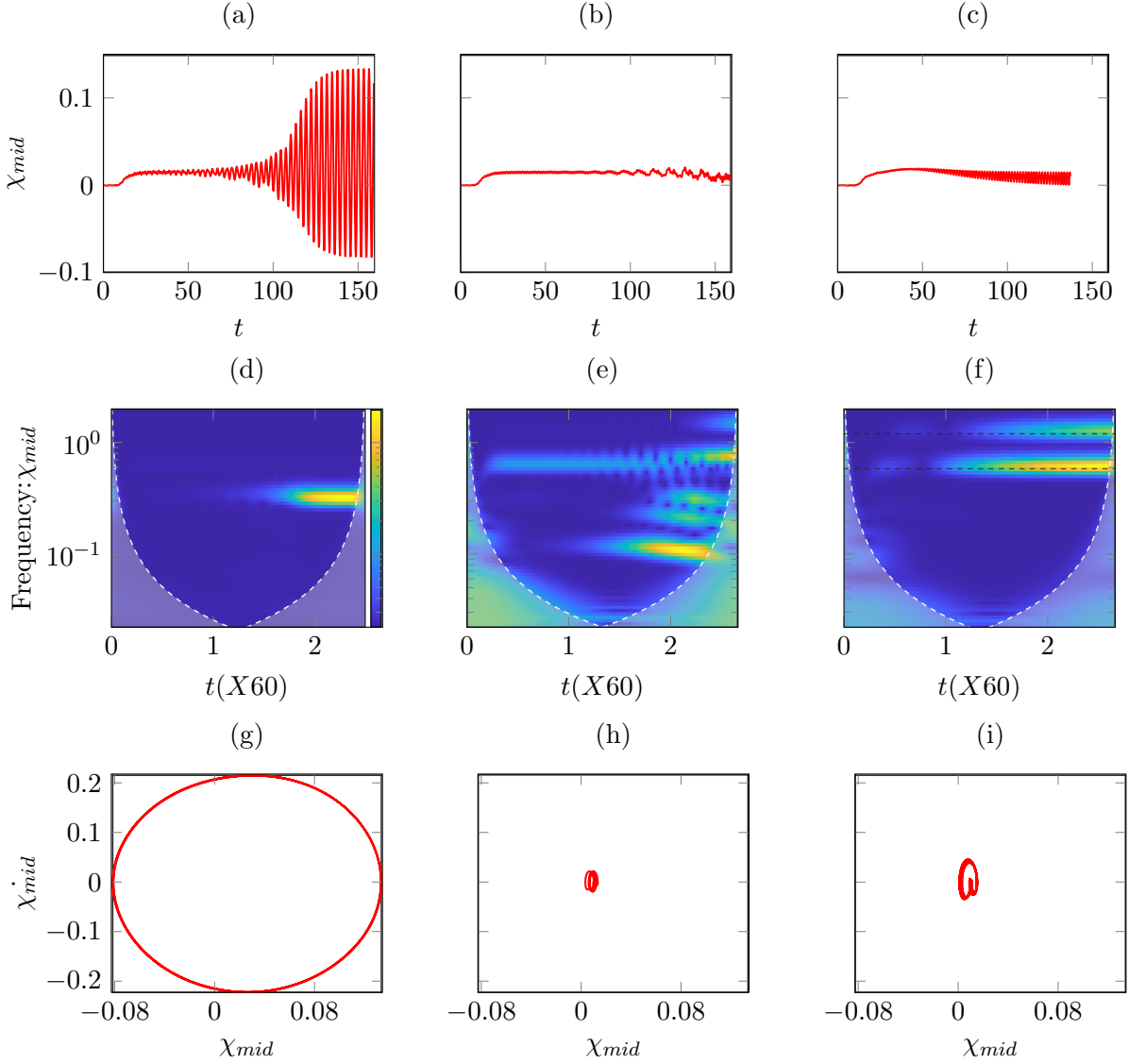


\begin{subfigure}[b]{0.35\textwidth}
\input{figures/mode2figures/Fig19a.tex}  
\end{subfigure}
\begin{subfigure}[b]{0.32\textwidth}
\input{figures/mode2figures/Fig19b.tex}  
\end{subfigure}
\begin{subfigure}[b]{0.31\textwidth}
\input{figures/mode2figures/Fig19c.tex}  
\end{subfigure}

\begin{subfigure}[b]{0.35\textwidth}
\input{figures/mode2figures/Fig19d.tex}  
\end{subfigure}
\begin{subfigure}[b]{0.32\textwidth}
\input{figures/mode2figures/Fig19e.tex}  
\end{subfigure}
\begin{subfigure}[b]{0.31\textwidth}
\input{figures/mode2figures/Fig19f.tex}  
\end{subfigure}

\begin{subfigure}[b]{0.35\textwidth}
\input{figures/mode2figures/Fig19g_pportrait.tex}  
\end{subfigure}
\begin{subfigure}[b]{0.32\textwidth}
\input{figures/mode2figures/Fig19h_pportrait.tex}  
\end{subfigure}
\begin{subfigure}[b]{0.31\textwidth}
\input{figures/mode2figures/Fig19i_pportrait.tex}  
\end{subfigure}
\caption{Time trace (a-c) and wavelet spectrogram (d-f) of the $\chi_{mid}$ signal, along with the phase portrait (g-i) of long-time $\chi_{mid}$ oscillations. Each column
corresponds to a different mass value $m$, as: $m = 2.76$ (a,d,g), $m = 4.14$ (b,e,h), and
$m = 5.5215$ (c,f,i). The wavelet spectrograms are normalized from $[0,1]$, and the entries are not marked in (d-f) for cleanness.}
\label{fig:modetransmidpt}
\end{figure}

We now consider the dynamics associated with $m \in [4.14, 6.62]$ (the leftmost portion of regime (C) from figure \ref{fig:spectral-projection}, corresponding to dynamics where the structural response is governed by mode two behavior). Figure \ref{fig:modetransmidpt} shows information about the time evolution in $\chi_{mid}$ for three representative cases: $m=2.76, 4.14, 5.52$ (the former case is one of the largest mass values that yields mode-one dominated behavior, and will be used throughout this section as a point of comparison for the higher-mode dynamics that arise at larger mass values). The $\chi_{mid}$ time trace for $m=2.76$ shows low-amplitude oscillations until $t\approx75$ and transitions to high-amplitude oscillations by $t\approx120$ (see figure \ref{fig:modetransmidpt}(a), and note the consistency with the lock-on behavior identified in the mode one regime from the prior section). The phase portrait associated with the plate's midpoint is elliptical in shape (see figure \ref{fig:modetransmidpt}(g)) implying that the plate shows mode one type oscillations. The wavelet spectrogram associated with the $\chi_{mid}$ signal shows that the deformation behavior is focused near a frequency of $0.3$ after $t=100$. Until about $t\approx75$, a faintly visible mark appears near $f_s=0.62$. The shift in frequency from $f_s$ to roughly $0.3$ is a consequence of the FSI system locking onto the fundamental mode 1 vibration dynamics, as discussed previously. 

For $m=4.14$, we notice distinctions in the dynamics when compared to $m=2.76$. 
Firstly, the $\dot{\chi}_{mid}$ versus $\chi_{mid}$ phase portrait is much smaller in surface area (although still vaguely circular in shape) when compared to $m=2.76$ (see figure \ref{fig:modetransmidpt}(h)). The amplitude of $\chi_{mid}$ oscillations is only about $0.023$ as opposed to the large magnitudes (near $0.1$) seen in $m=2.76$. The spectrogram of the $\chi_{mid}$ dynamics shows moderate signatures at $f_s$ until $t\approx 100$. After $t\approx120$, prominent dynamic content is visible at both $f_s$ and $f\approx 0.1$ (and triadically consistent frequencies). The signature at $f_s$ gradually increases in energy with time. For $t>100$, the $\chi_{mid}$ signal shows quasi-steady oscillations that involve the baseline shedding frequency $f_s$. This interplay between structural responses at very low frequency ($f\approx0.1$) and the vortex shedding frequency associated with the rigid plate ($f\approx f_s$) are consistent with the result from
 figure \ref{fig:spectral-projection}(c), which demonstrated that $m=4.14$ contains both mode one and mode two shaped responses in its plate oscillations. This duality suggests a transition in parameter space (in this case, increasing mass value) where the decreasing structural frequency is becoming too low for the inherent vortex-shedding behavior to lock onto. At this intermediate mass value, signatures of the flow at this low frequency persist in the structural vibrations. At higher mass values, to be discussed next, the flow is unable to interact with the lower structural mode. For these higher mass values, fluid-structure interplay occurs between the underlying vortex-shedding behavior and higher structural modes, all near $f\approx f_s$.

The $\chi_{mid}$ time trace for $m=5.52$ shows high-frequency oscillations with  a modulated low-frequency component. From the spectrogram, energetic content appears near $f_s$ at $t\approx50$ and stays until the end of the sampled time. At about $t\approx 80$ another strong signature is seen near $2f_s$, a harmonic of the baseline vortex-shedding frequency. The knot-like shape of the associated $\chi_{mid}$ phase portrait (figure \ref{fig:modetransmidpt}(i))  is reflective of the associated quasi-steady structural oscillations. We recall that $m=5.52$ predominantly shows mode two shaped oscillations in the long-time limit (see figure \ref{fig:spectral-projection}(c)). Of particular interest is the growing strength in oscillation energy near the baseline vortex-shedding harmonic, $2f_s$ (figure \ref{fig:modetransmidpt}(f)). We next probe the detailed snapshots to identify how this harmonic structural content is generated and sustained as part of the mode-two-shaped plate oscillations. 

\begin{figure}
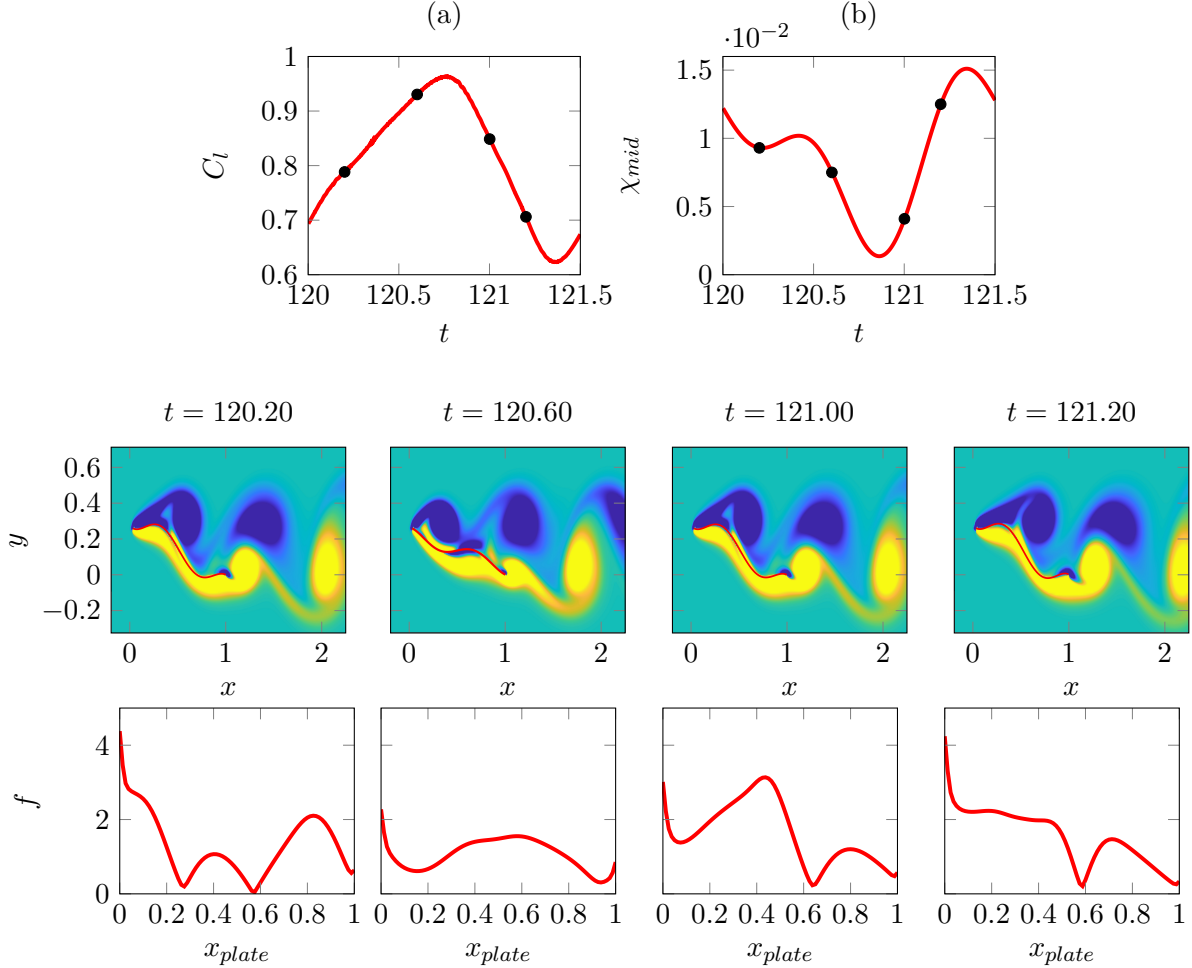

 \begin{subfigure}[b]{0.33\textwidth}
\input{figures/mode2figures/clcyclecomp_m5.52.tex}
\end{subfigure}
\begin{subfigure}[b]{0.32\textwidth}
\input{figures/mode2figures/midptcyclecomp_m5.52.tex}
\end{subfigure}

\begin{subfigure}[b]{0.29\textwidth}
\input{figures/mode2figures/Fig20a.tex}
\end{subfigure}
\begin{subfigure}[b]{0.22\textwidth}
\input{figures/mode2figures/Fig20b.tex}
\end{subfigure}
\begin{subfigure}[b]{0.22\textwidth}
\input{figures/mode2figures/Fig20c.tex}
\end{subfigure}
\begin{subfigure}[b]{0.22\textwidth}
\input{figures/mode2figures/Fig20d.tex}
\end{subfigure}

\begin{subfigure}[b]{0.28\textwidth}
\input{figures/mode2figures/Fig20e.tex}
\end{subfigure}
\begin{subfigure}[b]{0.22\textwidth}
\input{figures/mode2figures/Fig20f.tex}
\end{subfigure}
\begin{subfigure}[b]{0.22\textwidth}
\input{figures/mode2figures/Fig20g.tex}
\end{subfigure}
\begin{subfigure}[b]{0.22\textwidth}
\input{figures/mode2figures/Fig20h.tex}
\end{subfigure}
\caption{Top row (a,b): Magnified view of a representative portion from the $C_l$ and $\chi_{mid}$ time trace for $m=5.52$, Middle row: Vorticity snapshots at marked time instances in the time series, for $m=5.52$.
Bottom row: Magnitude of the resultant
surface stress $f$ distribution along the length of the plate, at each instance}
\label{fig:mode2snapshot}
\end{figure}

 Figure \ref{fig:mode2snapshot}(a,b) show a representative period in the $C_l$ and $\chi_{mid}$ time trace associated with $m=5.52$. Vorticity snapshots and resultant surface stress profiles are shown for specified time instances  (see the black markers in figure \ref{fig:mode2snapshot}(a,b), and corresponding subfigures in rows two and three) within the period. Although both the $C_l$ and $\chi_{mid}$ time trace is aperiodic, a representative cycle from the time trace is selected to better analyze how the lift variations coexist with mode two shaped plate oscillations.  
 
  The vorticity contour at $t=120.20$ shows that the portion of the LEV near the plate's fore half develops a negative curvature to match the negative camber associated with the fore-half of the plate. It can also be seen that the portion of the LEV that exists near the downstream half of the plate conforms to the positive plate curvature. From $t=120.20$ to $t=120.60$, figure \ref{fig:mode2snapshot}(a) shows that the lift on the plate increases from $C_l\approx0.8$ to $C_l\approx0.9$. During this interval, the plate's upstream half moves upwards while the downstream half shows a downward trajectory (see vorticity snapshots in figure \ref{fig:mode2snapshot}). Figure \ref{fig:mode2snapshot}(b) shows that the plate's midpoint first moves upwards before moving to a slightly more downward position. At $t=120.60$, the portion of the LEV near the downstream half of the plate has slightly advected downstream, while the LEV near the front has grown slightly larger in size compared to at $t=120.20$. The surface stress profiles are also noted to be different from $t=120.20$, as shown in figure \ref{fig:mode2snapshot}(bottom row). While the surface stress distribution at $t=120.20$ shows two prominent peaks near $x_{plate}\approx0.4, 0.8$, the distribution at $t=120.60$ does not show prominent peaks near the plate's downstream half, possibly indicating a lower signature of the LEV on the surface stress distribution. 
  
  From $t=120.60$ to $t=121.00$, the lift on the plate is seen to drop from $Cl_\approx0.92$ to $C_l\approx0.85$ while the plate's midpoint shows a net downward motion. The upstream half of the plate moves upwards to a position of positive camber while the downstream half shows a downward motion to a negative cambered position. The portion of the LEV that was present near the plate's downstream half at $t=120.60$ can be seen to have advected further downstream to $x\approx1.5$. The LEV that is present near the upstream half of the plate at $t=120.60$ has morphed in shape to accommodate the positive camber of the upstream half of the plate. The LEV has also slightly advected to a downstream position when compared to $t=120.60$. The surface stress distribution at $t=121.00$ shows a prominent peak near $x_{plate}\approx0.4$, indicating that the a portion of the LEV is close to the plate's upstream half. From $t=121.00$ to $t=121.20$, the LEV close to the plate continues its advection downstream, and is associated with a drop in lift (see figure \ref{fig:mode2snapshot}(a)). The plate's midpoint shows an upward trajectory from $\chi_{mid}\approx0.004$ to $\chi_{mid}\approx0.012$, while the plate's upstream half continues its upward motion. The surface stress distribution at $t=121.20$ shows a flat profile until about $x_{plate}\approx0.6$, indicating the absence of an LEV signature. By $t=121.20$, a new  TEV has fully formed near the plate's trailing edge.

This representative cycle of plate-vortex interplay repeats over a timescale commensurate with $f_s$ (see figure \ref{fig:mode2snapshot}(a)), where  vortex shedding and mode two structural oscillations interact as described above. It is interesting that while the mode two structural oscillations introduce a new wavelength, half the length of the plate over which vortex shedding typically occurs, this new length scale does not affect the global vortex-shedding formation. That is, rather than introduce new vortex structures near the midpoint associated with the higher-mode structural vibrations, the structural dynamics only alter the underlying LEV-TEV process inherent to the rigid plate system, over the same full-plate length scale and timescale of $1/f_s$. That said, the mode two motion of the plate introduces fine scale vortex flow features near the midpoint, which are eventually absorbed into the larger scale LEV and TEV processes but which could be introducing FSI dynamics at $2f_s$. This pathway towards a higher frequency harmonic could be a trigger towards nonlinear interplay and eventual aperiodicity in the lift and midpoint displacement signals, which will be discussed below. Each of the cycles, of which a representative case was just described, is modulated by low frequency behavior commensurate with the structural mode one natural frequency (this modulation is visually evident in figure \ref{fig:mode2snapshot}(b)). That is, the plate's mode one component of motion provides a quasi-steady modulation to the coupled FSI dynamics.

\begin{figure}
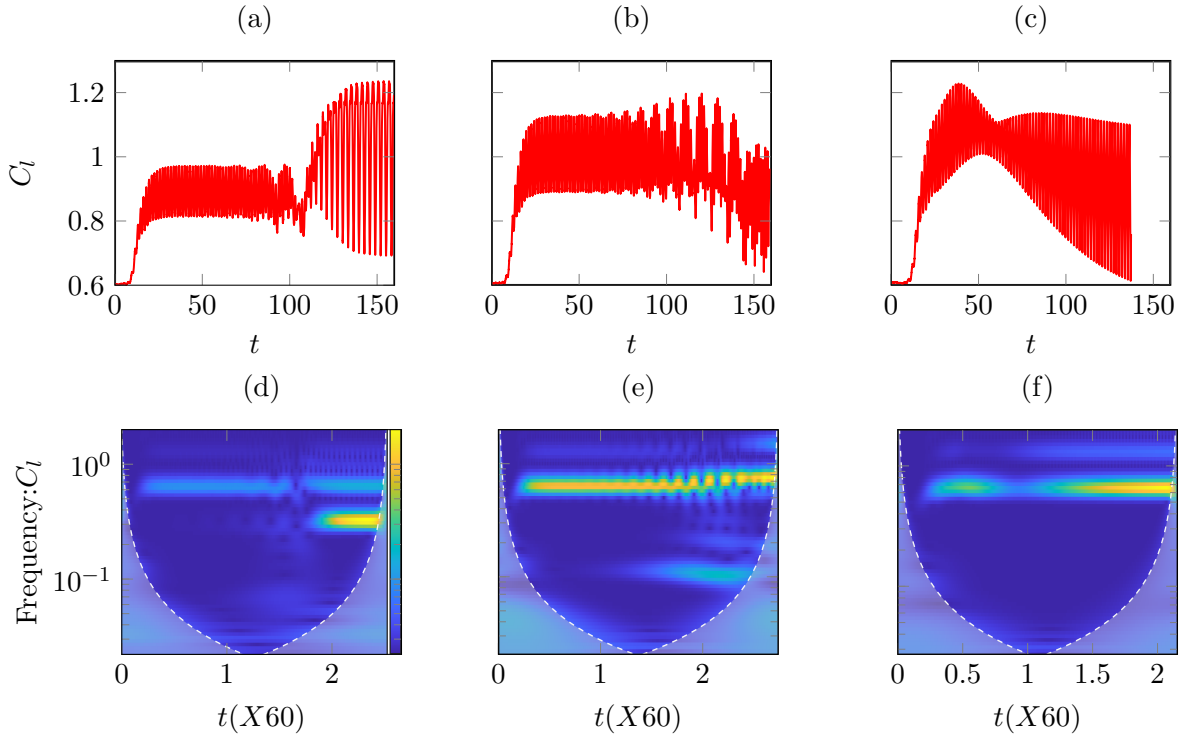


\begin{subfigure}[b]{0.35\textwidth}
\input{figures/mode2figures/Fig21a.tex}  
\end{subfigure}
\begin{subfigure}[b]{0.32\textwidth}
\input{figures/mode2figures/Fig21b.tex}  
\end{subfigure}
\begin{subfigure}[b]{0.31\textwidth}
\input{figures/mode2figures/Fig21c.tex}  
\end{subfigure}

\begin{subfigure}[b]{0.35\textwidth}
\input{figures/mode2figures/Fig21d.tex}  
\end{subfigure}
\begin{subfigure}[b]{0.32\textwidth}
\input{figures/mode2figures/Fig21e.tex}  
\end{subfigure}
\begin{subfigure}[b]{0.31\textwidth}
\input{figures/mode2figures/Fig21f.tex}  
\end{subfigure}

\caption{Time trace plots (a-c) and wavelet spectrogram (d-f) of $C_l$. Each column corresponds
to a different mass value $m$, as: $m = 2.76$ (a,d), $m = 4.14$ (b,e), and $m = 5.52$ (c,f). The wavelet spectrograms are normalized from $[0,1]$.}
\label{fig:modetranscl}
\end{figure}

 To clarify how the aperiodic plate motion and their ensuing changes to key vortex structures affect aerodynamic lift, figure \ref{fig:modetranscl} shows information about the evolution in $C_l$ for $m=2.76, 4.14, 5.52$. From the time trace, $m=2.76$ shows a shift from low amplitude to high-amplitude oscillations at $t\approx100$. The low-amplitude oscillations (until $t\approx100$) take place at a frequency near $f_s$ (see spectrogram in (d)) while the high-amplitude oscillations show a slightly lower frequency (see the dominant frequency content in  the spectrogram near $0.3$). The initial behavior near $f_s$ and subsequent lock-on to the first structural mode is consistent with what was described in detail in section \ref{sec:mode_1_detailed}. The $C_l$ time trace for $m=4.14$ shows oscillations near $f_s$ until $t\approx100$ (see time trace (b) and the energetic content near $f_s$ in the spectrogram (e)). After $t\approx110$, the time trace shows the emergence of low frequency modulations near a frequency of $0.1$ in the spectrogram. For $m=5.52$, the $C_l$ contains dominant content at the baseline vortex-shedding frequency of $f_s$, with visually evident but less prominent behavior at $2f_s$ (see faintly visible mark on spectrogram (f)). This frequency content evident from the spectogram is manifested in the time signal as an aperiodic waveform, possibly involving beating from these two frequencies. 

 Summarizing the behavior of the mode two regime, $m\in[4.14,6.62]$, the shift in the plate's oscillation mode is found to significantly impact the associated vortex-shedding process. This structure mode two regime yields coupled dynamics with dominant frequency content at $f_s$, and a subsequent appearance of its harmonic $2f_s$. We argue that this sequential triggering of the harmonic is due to the new length/timescale introduced by the mode two structural motion, and provides a pathway for the aperiodic behavior observed in the FSI system. 
 
 First, the proximity of the vortex-shedding frequency to the structural mode two natural frequency triggers mode two-dominated structural motion (c.f., figure \ref{fig:parametricspace}(b)). Note that this appearance of mode two behavior with increasing $m$ comes after a structure mode one regime, where with increasing $m$ the flow locks onto progressively lower frequencies associated with the structure mode one frequency. These lower frequencies are associated with a decreasing correlation in the LEV-$\chi_{mid}$ correlation (c.f., figure \ref{fig:mode1-phase}(c)). Along with the decreasing phase relationship between the vortex structures and the mode one plate oscillations, the increasing mass also yields a mode two natural frequency that is gradually closer to $f_s$. For sufficiently high $m$, these two effects trigger a bifurcation in the observed structural mode (and, by extension, the surrounding flow structures). 

The plate's mode two shaped oscillations trigger flow structures that are roughly the same size as half the length of the plate (commensurate with the spatial wavenumber of the plate's mode two shape). The vortex structures (see LEV formation and advection in the middle row in figure \ref{fig:mode2snapshot}), although spatially altered by the mode two oscillations at $2f_s$, still continue to form and advect predominantly at the baseline shedding frequency $f_s$. The timescales corresponding to the twice harmonic component (associated with the mode two oscillations), and the fundamental mode one frequency (associated with the mode one oscillations) continue to modulate the subsequent dynamic interplay.

 This behavior is in contrast to the cases where the FSI behavior was associated with structural mode one vibrations. In that setting, the structure and flow synchronize onto dynamics that evolve either almost near the baseline vortex shedding frequency $f_s$ (c.f., the fluid-induced mass regime of section \ref{sec:add_mass_detailed}) or the vacuum-scaled mode one natural frequency (c.f., the $m=2.76$ case and higher mass values within the structure mode one regime). Note also that the mode two regime, $m\in[4.14,6.62]$, contains intermediate cases such as that of $m=4.14$ where there is dominant frequency content associated with both the fully mode one and fully mode two regimes.

\begin{figure}
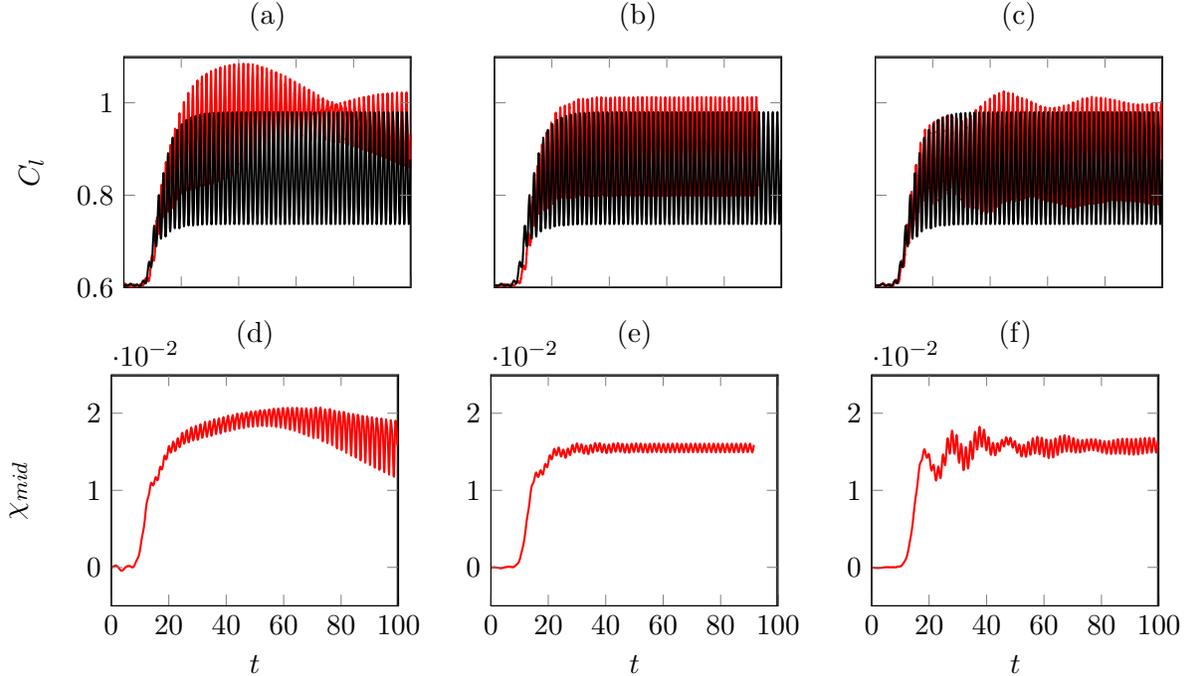

\begin{subfigure}[b]{0.36\textwidth}
\input{figures/mode3figures/m6.62_lift.tex}   
\end{subfigure}
\begin{subfigure}[b]{0.3\textwidth}
\input{figures/mode3figures/m11.1004_lift.tex}  
\end{subfigure}
\begin{subfigure}[b]{0.3\textwidth}
\input{figures/mode3figures/m33.31_lift.tex}  
\end{subfigure}

\begin{subfigure}[b]{0.36\textwidth}
\input{figures/mode3figures/m6.62_midpt.tex}   
\end{subfigure}
\begin{subfigure}[b]{0.3\textwidth}
\input{figures/mode3figures/m11.1004_midpt.tex}  
\end{subfigure}
\begin{subfigure}[b]{0.3\textwidth}
\input{figures/mode3figures/m33.31_midpt.tex}  
\end{subfigure}

\caption{Time trace plots of the $C_l$ (a-c) and $\chi_{mid}$ (d-f) signals. Each column corresponds
to a different mass value $m$, as: $m = 6.62$ (a,d), $m = 11.10$ (b,e), and $m = 33.32$ (c,f). Black curve indicates the $C_l$ time trace for rigid-baseline.}
\label{fig:highermodefig}
\end{figure}

We last consider the cases where even higher structural modes are triggered, $m>6.62$. These cases receive less focus because of their minimal impact on the flow and associated aerodynamics relative to the rigid case (c.f., figure \ref{fig:highermodefig}). Figure \ref{fig:highermodefig} shows the $C_l$ and $\chi_{mid}$ time traces for $m=11.10$ and $m=33.32$ ((b),(c),(e),(f)) alongside the mode two $m=6.62$ ((a),(d), provided for reference). From the  $\chi_{mid}$ time trace, $m=11.10$ leaves its position of zero-mean deflection at $t\approx10$ and settles into low-amplitude oscillations about a non-zero mean deflection by $t\approx40$. The corresponding time trace for $C_l$ qualitatively resembles the evolution in the baseline case (see black time trace in (b)) albeit with a slightly higher $\overline{C_l}$. The $\chi_{mid}$ time trace does not appear to modulate or affect the $C_l$ dynamics, notably in contrast to $m=6.62$ where $\chi_{mid}$ provides a low-frequency modulation to $C_l$. For $m=33.32$, figure \ref{fig:highermodefig}(f) shows that the $\chi_{mid}$ time trace is qualitatively similar to the one seen in $m=11.11$. However, the time trace does show the existence of low-frequency modulations until $t\approx60$. These modulations decay over $t\in[60,80]$, and the oscillations thereafter are of low amplitude and comprised of frequency $f_s$. The plate oscillations after $t\approx60$ are mode three in shape, in contrast to the mode two oscillations seen for $m=6.62$ (see figure \ref{fig:spectral-projection} (c)). The $C_l$ time trace for $m=33.32$ shows an initial modulation that is driven by low-frequency content (see \ref{fig:highermodefig}(c)) until $t\approx80$. The modulation appears to decay after $t\approx80$ and the oscillations thereafter resemble those of the baseline (rigid) case.

\section{Conclusion}
In this manuscript, we have investigated the fluid-structure interaction (FSI) between a flexible flat plate and an unsteady aerodynamic flow using high-fidelity 2D numerical simulations. For the chosen aerodynamic parameters of Re=$500$ and a post-stall angle of attack (15$^{\circ}$), the baseline flow for the rigid flat plate was separated, with vortex shedding taking place at a characteristic frequency $f_s$. For our study with the flexible flat plate, we selected our parameters such that the fixed stiffness $k=0.0265$ provided a low amplitude of mean-camber, allowing a focus of the effect of plate dynamics on the ensuing FSI and resulting aerodynamic lift. For plate oscillations about this fixed mean camber, we probed the role of mass $m\in(0.0007,33.31)$ in dictating the vortex-plate interplay. Our selection of $m$ was based on carefully aligning the baseline vortex-shedding frequency ($f_s$) with the fundamental frequencies of the plate.   

We performed our numerical simulations by starting from a formal steady unstable base state of flow-structure system, where the plate was undeformed with a pre-stress applied so that the plate would remain undeformed under the steady base flow. For all masses considered, the early time dynamics from the the base state had nearly identical frequency content. The results suggested that the departure from the base state was instigated by a ``flow-driven" instability associated with the bluffness of the plate. That is, for early time it is the geometry of the plate that triggers flow separation and early vortex shedding, and subsequent flow-structure interplay is a response to that stimulus. In the long-time limit, significant variations were identified in the vortex-plate interplay that led to the identification of demarcated mechanistic regimes. These regimes were classified based on the structural mode that was induced during the flow-structure interplay, with structural behavior dominated by mode one (regimes A and B), and modes two and three (regime C) observed. Our focus  was to identify the mechanisms that triggered the onset of the regimes, to clarify how the dominant structural timescales interacted with the underlying vortex-shedding content (nominally at $f_s$ in the rigid case), and how these effects conspired to affect aerodynamic lift.

For low $m\in[0.0004, 0.26)$ (regime A), the $\overline{C_l} (\approx0.948)$ and $\chi_{mid} (\approx0.019)$ were approximately found to be the same for all the $m$. The peak-to-peak amplitude in $C_l$ and $\chi_{mid}$ were found to be roughly constant across all $m$, and the frequency of oscillations closely resembled the baseline vortex-shedding frequency $f_s$. The limit-cycle oscillations for a representative case in the regime showed several qualitative distinctions from the baseline rigid case---prominently, a higher magnitude in the $C_l$ oscillations and a differing LEV-TEV phase relationship. The plate oscillations predominantly showed a mode one shape in its response and maintained an in-phase relation with the LEV's formation and advection processes. Across the regime, the plate-vortex interplay was found to be dictated by the fluid-induced mass, which was computed to be much larger than the structural mass.        

For $m\in[0.26,2.76)$ (regime B), the $\overline{C_l}$ showed a consistently increasing trend with increasing mass $m$, with $\overline{C_l}=1.05$ for $m=2.76$. The $\overline{\chi_{mid}}$ and peak-to-peak $\chi_{mid}$ excursions showed a roughly increasing trend while the peak-to-peak $C_l$ excursions demonstrated an increasing and then decreasing trend. The plate oscillations in the long-time limit showed a mode one-dominated response. For $m\in[0.26,2.76)$, the phase relation between the plate motion and the LEV circulation strength  ($\Gamma_{LEV}$) reduced from approximately $0.9$ to $0$. The frequency of plate oscillations aligned closely with the first fundamental in-vacuo frequency of the plate. The limit-cycle interplay  strongly indicated a resonant ``lock-on" mechanism. That is, the time evolution for any $m$ within the regime showed an extended transient phase where the plate was initially forced by fluctuations from the baseline vortex shedding near $f_s$. The plate oscillations were found to eventually shift to a  lower frequency with increasing $m$, with large-amplitude mode one oscillations aligned with the first natural frequency of the plate. For larger mass values the structure mass dominated the fluid-induced mass, so that $\nu_1\approx \nu^{vacuum}_1$. Consistent with this outcome, the dominant frequency associated with the $\chi_{mid}$ and $C_l$ dynamics showed a continuously decreasing trend with increasing $m$, increasingly aligned with $\nu^{vacuum}_1$.

For $m\in[2.76, 33.32]$ (regime C), $\overline{C_l}\approx0.88$ throughout the regime, only slightly greater in magnitude than $\overline{C_l}_{rigid}$. The peak-to-peak excursions in both $C_l$ and $\chi_{mid}$ were modest in comparison to regime (B). While the frequency content associated with $C_l$ showed
values similar to the baseline rigid case, a significant spread was found in the frequencies of the dynamics associated with $\chi_{mid}$. The plate oscillations showed a mode two-dominated response for $m\in[4.14,6.62]$ and a mode three-dominated response for $m\in[11.10, 33.32]$. The mode-two cases were associated with quasi-steady dynamics in both $C_l$ and $\chi_{mid}$ that never settled onto a limit-cycle attractor. In the long-time limit, the mode two deformation dynamics significantly modified the LEV formation and advection processes, leading to excitation of the $2f_s$ harmonic. For higher mass values, mode three oscillations arose and were found to be too small in amplitude to have a significant impact on $C_l$.

\section{Acknowledgement}
The authors gratefully acknowledge support from the Air Force Office of Scientific Research under Award No. FA9550-21-1-0182.
\section{Appendix}
\label{sec:App}

\subsection{Overview of regimes for different stiffness values}
\begin{figure}[h!]
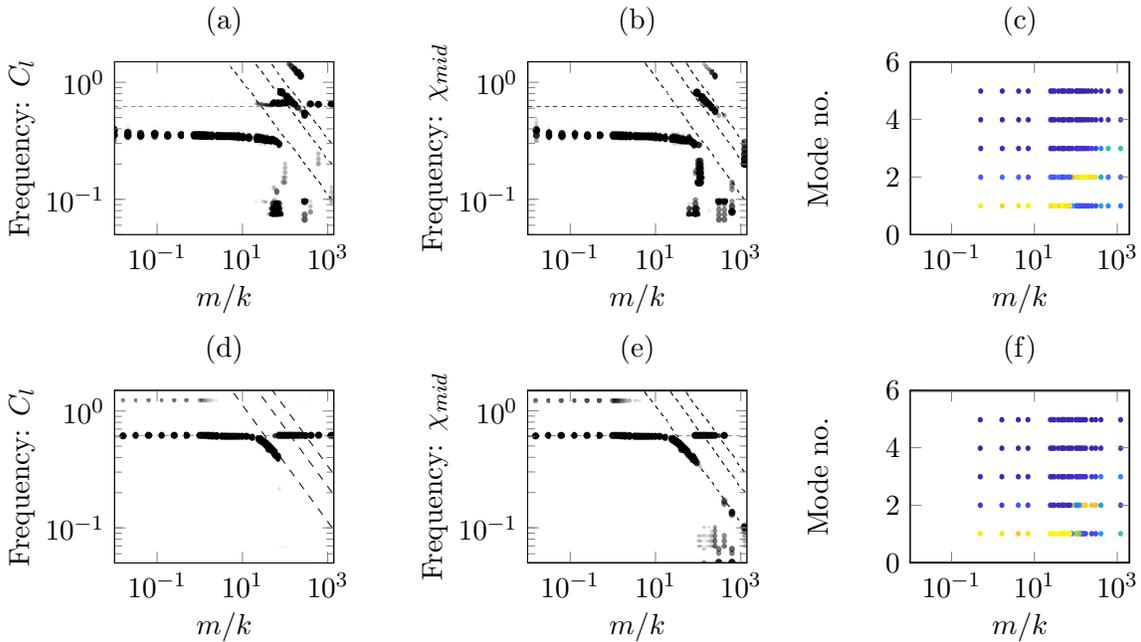

\centering 
\begin{subfigure}[b]{0.36\textwidth}
\input{figures/appendixfigures/Fig23a_00}\end{subfigure}
\begin{subfigure}[b]{0.3\textwidth}
\input{figures/appendixfigures/Fig23a_01} \end{subfigure}
\begin{subfigure}[b]{0.3\textwidth}
\input{figures/appendixfigures/Fig23a_02} \end{subfigure}

\begin{subfigure}[b]{0.36\textwidth}
\input{figures/appendixfigures/Fig23a_10}\end{subfigure}
\begin{subfigure}[b]{0.3\textwidth}
\input{figures/appendixfigures/Fig23a_11} \end{subfigure}
\begin{subfigure}[b]{0.3\textwidth}
\input{figures/appendixfigures/Fig23a_12} \end{subfigure}
\caption{Overview plots of behavioral regimes for $k=0.008745$ (a-c) and $k=0.106$ (d-f). Power spectral density (PSD) associated with the long-time $C_l$ (a,d) and $\chi_{mid}$ (b,e). Projection of the structural dynamics onto the various structural modes. The colorbar scale is the same as in figure \ref{fig:spectral-projection}(c)}.
\label{fig:appendixfig}
\end{figure}

\begin{figure}  
\begin{subfigure}[b]{0.3\textwidth}
\input{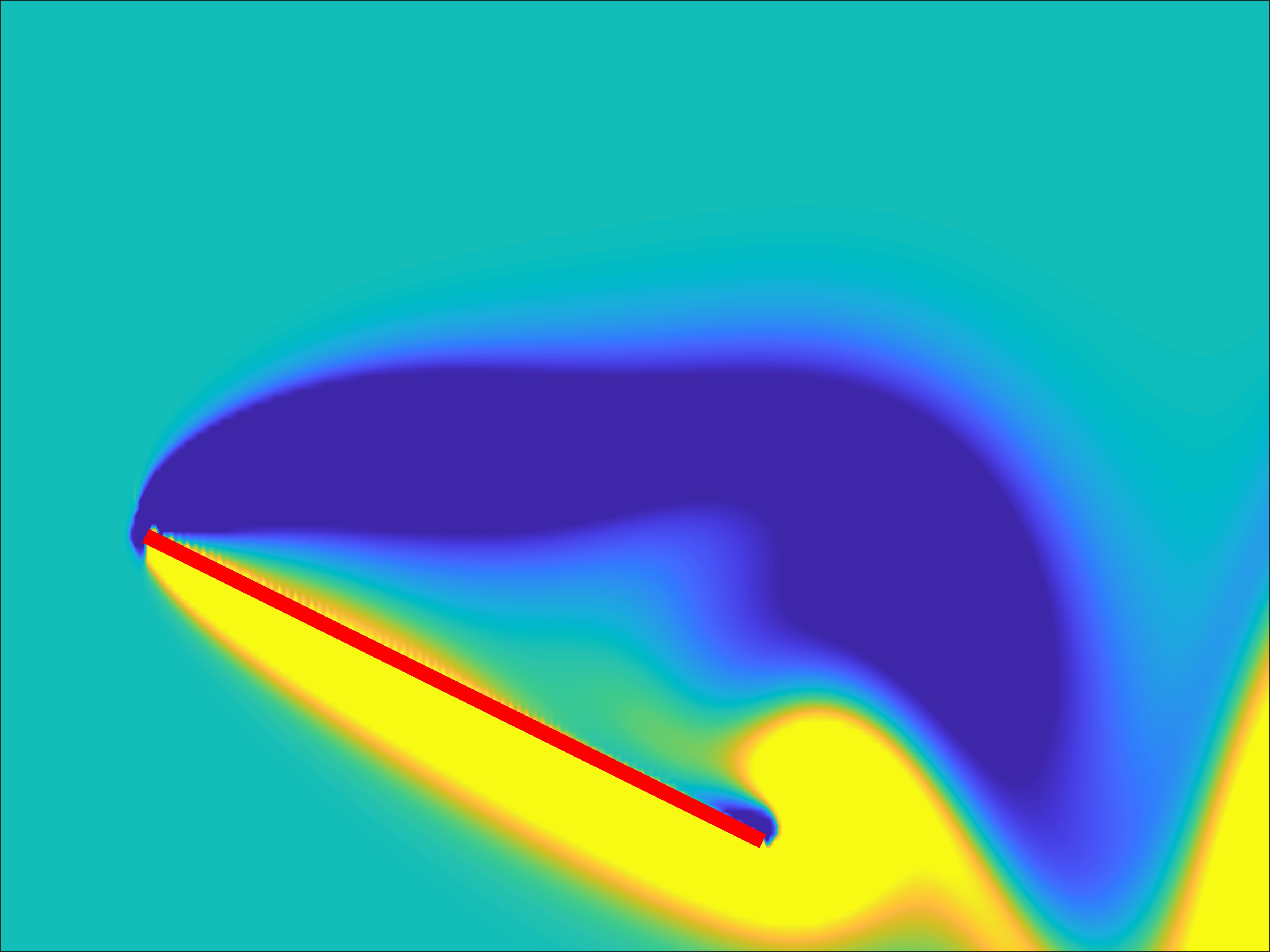} 
\end{subfigure}
\caption{Integration contours used in the quantification of LEV (black) and TEV (purple) circulation strengths.}
\label{fig:strength_contour}
\end{figure}

To augment the focus of the article on the chosen value of $k=0.02$, we provide here a brief summary of the dynamics for a lower ($k=0.106$) and a higher ($k=0.0087$) mean camber. We leave detailed analysis of these different stiffness values to future work, and instead use overview plots to demonstrate regime commonalities across these three cases. We will also highlight potential differences where appropriate. We note that the range of $m/k$ values explored is kept constant across the three mean cambers: the goal is to study the relative positioning of vortex shedding frequency with respect to the fundamental frequencies. That is, the values of $m$ considered is largest for $k=0.106$ and lowest for $k=0.008745$. 

Figure \ref{fig:appendixfig} is an analog of figure \ref{fig:spectral-projection} for the two separate $k$ values considered in this appendix. This figure provides an overview of the spectral content for the $C_l$ and the $\chi_{mid}$ signal (see the main text surrounding figure \ref{fig:spectral-projection} for full details). To enable comparison between the different stiffness values, the horizontal axis is re-scaled as $m/k$ and a discussion of figure \ref{fig:spectral-projection} is incorporated into the text of this appendix to fit within the context of the re-scaled axis.

Figure \ref{fig:appendixfig}(a) shows the spectral content associated with the $C_l$ signal for $k=0.008$. Until $m/k\approx12.5$, the markers show that the dominant frequency is near frequency $\approx 0.35$.  From $m/k\in(25,50)$ the dominant markers indicate a slight decrease in the dominant frequency content. At $m/k\approx50$, the dominant frequency is $\approx0.32$. For $m\in(50,83.33)$, markers of significant size and color are present near both frequency values of $0.08$ and $f_s$, respectively. For $m/k\in(83,125)$, the dominant lift dynamics are at a frequency of $f_s$. For $m/k\in(125,312)$, the markers align closely along the dashed line representing the second natural frequency of the plate (also near $f_s$). The markers over this $m/k$ range also demonstrate significant content near $f\approx2f_s$. For $m/k>312$, the markers demonstrate that the lift response is singularly dominated near $f_s$. Figure \ref{fig:appendixfig}(d) shows the analog for the lower mean camber (higher stiffness), $k=0.1$. For $m/k<16.66$, the markers indicate a lift response dominated near the frequency $f_s$. The dominant markers show a linearly decreasing trend along the first fundamental frequency from $m/k\in(16.67,73.5)$. At $m/k\gtrsim59.5$, the lift response again acquires dominant signatures near the vortex-shedding frequency, $f_s$. 

Figures \ref{fig:appendixfig}(b,e) show the analog to figures \ref{fig:appendixfig}(a,d) for the spectral content associated with the $\chi_{mid}$ signal. These figures demonstrate qualitatively similar regime changes but with detailed differences in where these changes occur, and what frequencies are associated with them, detailed below. 
In figure \ref{fig:appendixfig}(b), for $m/k<10$, the markers indicate a dominant frequency content of $\approx0.35$, and show a slight dip to $\approx 0.3$ at $m/k\approx83.32$. For $m/k\in(96,250)$ the markers align closely along the second fundamental mode, with appreciable density of markers also occupying lower harmonics. For $k=0.1$, figure \ref{fig:appendixfig}(e) indicates that markers occupy a frequency of $\approx f_s$ until about $m/k\approx16.67$. From $m/k\in(25,96)$, the markers align along the first natural frequency indicating a linearly decreasing dominant frequency. For $m/k>96$, the markers occupy frequency $\approx f_s$.

Figure \ref{fig:appendixfig}(c) shows the scaled projection coefficient of the plate oscillations, for $k=0.008$. For $m/k<73.35$, the maximum energy (bright yellow) is seen on the mode one markers. At $m/k\approx83.32$, a transition is seen wherein the mode two markers indicate the largest projection. The mode two markers continue to show the dominant value until $m/k\approx312.6$. For $m/k\in[ 628, 1256]$, a significant proportion of the projection coefficient appears near mode three, although the large mean displacement also yields a large mode 1 contribution. The trends in the projection coefficient for $k=0.1$ is shown in figure \ref{fig:appendixfig}(f). Until $m/k\approx125$, the dominant mode is shown to be mode one. For $m/k\in(138.89,312.62)$ the markers indicate that mode two is the dominant shape in the plate oscillations. For high $m/k\approx1256$, a proportion of the coefficient ($\approx 30\%$) indicates a strong mode three shape in the oscillations.  

Broadly, these plots indicate similar regimes of behavior to those detailed in the main article with increasing $m$: first (regime A), a fluid-induced mass regime where the dynamics are at a near constant frequency near $f_s$. Second (regime B), a regime where the dominant mode one response induces lock-on of the flow onto this mode one natural frequency that scales progressively as the vacuum-scaled natural frequency with increasing mass. Third (regime C), a break from this mode one-scaled natural frequency to FSI dynamics associated with higher structural modes. 
Moreover, the $m/k$  values at which these regime transitions occur is markedly similar across $m$ values that vary by orders of magnitude.
These similarities  demonstrate the importance of the natural frequency in setting the overall FSI dynamics.

At the same time, there are important differences in the details across these stiffness values, which we summarize below.
The spectral content ($C_l, \chi_{mid}$) for $k=0.0265$ (considered in the main manuscript in figure \ref{fig:spectral-projection}(a, b)) shows an appreciable decrease in the dominant frequency (from $f\approx0.53$ down to $f\approx0.3$) as $m/k$ is increased across $m/k\in[8.3,104.15]$. This regime A dominant frequency is consistent with what was observed for $k=0.106$, but distinct from the $k=0.008745$ case. The differences for the lower stiffness suggest that the larger mean excursion of the beam drive a change in vortex-shedding content, and thereby in the structural dynamics. For $k=0.0265$, the markers corresponding to the $C_l$ spectral content  align with the second natural frequency for $m/k\in(178,250)$ and $f_s$ for $m/k>250$. While the corresponding markers for $k=0.008$ do show a similar alignment, no markers are found to explicitly align along the second mode for $k=0.1$. Instead, most markers occupy values near $f_s$ for $m/k>59.5$. This observation seems to suggest that the second mode structural dynamics are more prominently excited at lower stiffness values. The trends in the projection coefficient for $k=0.0265$ shows that the plate oscillations until $m/k\approx104.15$ are predominantly mode one and that a transition from mode one to mode two occurs in the range $m/k\in(124.9,138.86)$. A similar transition from mode one to mode two occurs over a slightly different  $m/k$ range for the other stiffness values considered ($m/k\in(73.35, 83.32)$ for $k=0.008$ and over $m/k\in(125-138.88)$ for $k=0.1$).

\subsection{Definition of LEV and TEV for this work}
 This article utilized definitions for computing the circulation strengths near the plate. To help explain the definitions, figure \ref{fig:strength_contour} shows a snapshot from the shedding cycle associated with the rigid-baseline case, at $t/T=0$.  The LEV and TEV strength is computed as

 \begin{equation*}
\Gamma_{LEV}=\int_{A_{black}}\gamma^{-}dA,
\end{equation*}
\begin{equation*}
\Gamma_{TEV}=\int_{A_{purple}}\gamma^{+}dA, 
\end{equation*}
 where $\Gamma$ is the circulation strength of the vortex. $\gamma^-$ and $\gamma^+$ are the magnitudes of the clockwise 
 and anti-clockwise vorticities associated with the LEV and TEV respectively. The integration on $\gamma^-/\gamma^+$ is performed on the area ($A_{black}/A_{purple}$) enclosed by the black/ red dashed lines respectively (see figure \ref{fig:strength_contour}). The black dashed lines are defined primarily using: (a) Horizontal line at $y=0.532$ that bounds the anti-clockwise vorticity content from top, thereby also leaving additional buffer for vortex motion due to plate displacement, (b) Dashed diagonal line along the plate's length from the leading-edge to the trailing edge (coordinates $(0.01,0.26)$ to $(1.0,0.0)$), (c) Two vertical lines at $x= 0.01, 1.09$ that horizontally bound the vorticity content relevant to the plate dynamics. The purple dashed lines are used to define the TEV strength, and are defined as a rectangular contour that surrounds the trailing edge of the plate ($[0.99, 1.18]\times [-0.08, 0.09]$).

\bibliography{references}

\end{document}